\documentclass[epj]{svjour}
\usepackage{graphics}
\usepackage{graphicx}

\newcommand{\be}{\begin{equation}}
\newcommand{\ee}[1]{\label{#1} \end{equation}}
\newcommand{\ba}{\begin{eqnarray}}
\newcommand{\ea}[1]{\label{#1} \end{eqnarray}}
\newcommand{\nl}{\nonumber \\}

\newcommand{\nbar}{\overline{n}}
\newcommand{\sbar}{\overline{s}}
\newcommand{\ebar}{\overline{e}}

\newcommand{\rint}[1]{{\int\!d^dr_{#1} \, }}
\newcommand{\pint}[1]{{\int\!\frac{d^dp_{#1}}{(2\pi\hbar)^d} \, }}
\newcommand{\pd}[2]{{\frac{\partial #1}{\partial #2}}}

\newcommand{\vs}{\vspace{3mm}}

\begin{document}
%
\title{Non-Extensive Approach to Quark Matter}
\author{Tam\'as S. Bir\'o\inst{1}, G\'abor Purcsel\inst{1}  \and K\'aroly \"Urm\"ossy\inst{1}} 
%
%
\institute{KFKI Research Institute for Particle and Nuclear Physics}
\date{Received: date / Revised version: date}
%
\abstract{
 We review the idea of generating non-extensive stationary distributions based on
 abstract composition rules of the subsystem energies, in particular the parton
 cascade method, using a Boltzmann equation with relativistic kinematics and
 modified two-body energy composition rules. The thermodynamical behavior of
 such model systems is investigated. As an application hadronic spectra with power-law
 tails are analyzed in the framework of a quark coalescence model.
} 
\maketitle
\section{Introduction}
\label{intro}

Power-law tailed distributions are abundant in Nature and in human technology ranging
from high energy particle spectra to fluctuations in stock markets or connectivity statistics
in the Internet. It would be natural to explain this abundance by a universal, statistical
limiting distribution since different causes result in similar outcomes. A more prestigious
attempt is to set such phenomena into a united framework of non-extensive thermodynamics,
based on certain generalizations of familiar basic formulas. In particular generalizations
of the Boltzmann -- Gibbs -- Shannon entropy formula were seeked as funding stones for such a
general treatment \cite{GE1,GE2,GE3,GE4,Fis59b,Van06a,OTHER-ENTROPIES,TSALLIS-ENTROPY}.

Several basic questions arise during this enterprise: among those the uniqueness of
equilibrium state and the entropy function describing irreversibility, the connection
between composition rules for basic thermodynamical quantities between two large
subsystems and the extensivity limit for a system with a large number of degrees of freedom,
and the very question that which microscopical mechanisms lead to such a distribution.
Is this a sign of non-equilibrium, of incomplete equilibrium or just of a new, generalized
kind of equilibrium? Applying and justifying a statistical, least thermodynamical approach
to high energy heavy ion collisions, as it is being central in the experimental quest
for quark matter, in particular requires clarification of the above questions.
Any inference to a thermal state and a physical temperature of the quark matter from
single particle spectra must connect the fit parameter measuring the spectral slope to
basic principles of thermodynamics.

In recent years we have been succeeding towards answers to the above problems.
After facing the fact that transverse momentum spectra fit well to a cut power-law
distribution towards much higher values than just the simple Gibbs-exponential,
a particular parton cascade approach was suggested in Ref.\cite{NEBE}
for generating these distributions. It has been observed that the stationary
distribution generated and maintained by a Boltzmann type equation is intimately
related to the energy composition rule used in two-particle encounters.
A simple modification of the kinetic energy addition rule among two partners,
which in high energy collisions is probably related to the relativistic kinematics,
leads to the observed result. A general treatment of abstract composition rules
is presented in Ref.\cite{EPLBiro}, where the non-extensivity property is related
to the deformation of addition rule and hence to the deformation of the classical 
Gibbs exponential.

The physics' question to begin with is the source of non-extensivity, especially
for the two most relevant quantities, energy and entropy. 
It is relatively easy to construct examples with non-extensive energy, whenever the
interaction retails a fractal structure in the phase space and therefore cannot be
neglected in the large volume -- large particle number limit, as it is traditional in
classical thermodynamics. It is much harder to understand non-extensive entropy,
however. In order to shed some light to possible mechanisms by which non-extensivity
in one-particle variables, like entropy and energy, can occur in physical systems, 
let us investigate a very particular case.

We assume that in an N-particle system there are two-particle correlations left
and seek for their relative contributions to total energy and entropy. For the sake
of demonstration we regard the following special form of the two-particle density:
\be
 \rho_{12} =  f(p_1) f(p_2) g(r_{12}),
\ee{TWO-PARTICLE-DENSITY}
which is factorizing in the momentum space via one-particle distribution functions,
but  is connected in the coordinate space via the pair-distribution function,
$g(r)$, of the relative coordinates.

The trace over states is determined via phase space integrals, normalized to satisfy
the following conditions in $d$ spatial and momentum dimensions:
\ba
 &\rint{} 1  & = V, \nl
 &\rint{} g(r) & = V_{{\rm eff}}, \nl
 &\pint{} f(p)  & = \nbar,
\ea{CONDITIONS}
with $V$ being the total volume, $V_{{\rm eff}}$ the available volume for a partner
of a given particle, and $\nbar$ the average (mean) density in the system. 
We normalize the integrals so that $\nbar V =N$ and $\nbar V_{{\rm eff}} = N-1$.

Under the above conditions this particular two-particle density is normalized to 
${\rm Tr} \rho_{12} = N(N-1)$.
The partial trace over the second particle leads to the familiar one-particle distribution
function used in kinetic theories:
\be
 \rho_1 = {\rm Tr}_{2} \left( \rho_{12} \right) = (N-1)  f(p_1).
\ee{PARTIAL_TRACE}

The total entropy of a correlated pair in matter,
\newline $-{\rm Tr}\left(\rho_{12}\ln\rho_{12}\right)/{\rm Tr}\left(\rho_{12}\right)$, 
is expressed by
\be
 S_2 = -\pint{1}\pint{2}\rint{1}\!\rint{2} \frac{ \rho_{12} \ln \rho_{12}}{N(N-1)}.
\ee{TOTAL-PAIR-ENTROPY}
In calculating this quantity two further individual integrals occur:
\ba
 \sbar &=& -\pint{} f(p) \ln f(p), \nl
 V_{{\rm info}} &=& - \rint{} g(r) \ln g(r).
\ea{INFO_INTEGRALS}
Using these notations one arrives at:
\be
 S_2 = 2 \, \frac{\sbar}{\nbar} + \frac{V_{{\rm info}}}{V_{{\rm eff}}}.
\ee{S2}
Generalizing the above expression valid for the two-particle density, $\rho_{12}$,
to an $N$-particle density, $\rho_{12\ldots N}$, factorized into $N(N-1)/2$ pair contributions
we obtain the following entropy per particle:
\be
 \frac{S_N}{N} =  \frac{\sbar}{\nbar} + \frac{\nbar}{2} \, V_{{\rm info}}.
\ee{SN}
The entropy of such a system is considered to be extensive, as long as the specific ratio remains finite in
the large particle number limit:
\be
 \lim_{N\rightarrow\infty}\limits \frac{S_N}{N} < \infty.
\ee{SLIMIT}
In this sense dangerous pair distributions are those, for which $V_{{\rm info}}$ increases
with $N$ at fixed mean density $\nbar$. In a  familiar piece of matter the pair distribution
function $g(r)$ approaches one at large distances, in these cases $V_{{\rm info}}$ is finite
and hence the entropy is extensive. In case of a quark gluon plasma, however, some infrared
magnetic modes remain non-perturbative and hence long range correlations remain. As a 
consequence $-g\ln g$ may not tend to zero fast enough and therefore the integral
$V_{{\rm info}}$ in eq.(\ref{INFO_INTEGRALS}) may increase as a function of $N$.
For example considering power-law type pair distribution functions, like
$g(r)=r^a/(1+r^b)$, the corresponding integrals up to a large radius, $R$ scale
like $V_{{\rm eff}} \sim R^{d+a-b}$ and like $V_{{\rm info}} \sim R^{d+a-b} \ln R$.
In this case the specific entropy for large $N$ becomes
\be
 \frac{S_N}{N} \longrightarrow \frac{\sbar}{\nbar} + {\rm const.} (a-b)N \ln N
\ee{LARGE_N_SPECIFIC}
with some unspecified constant. For $a \ne b$ this would lead to a non-extensive entropy.
For other possible sources of non-extensive entropy see Ref.\cite{ADDITIVE_NOT_LOG_ENTROPY}.

\begin{figure}
  \begin{center}
    \includegraphics[width=0.25\textwidth,angle=-90]{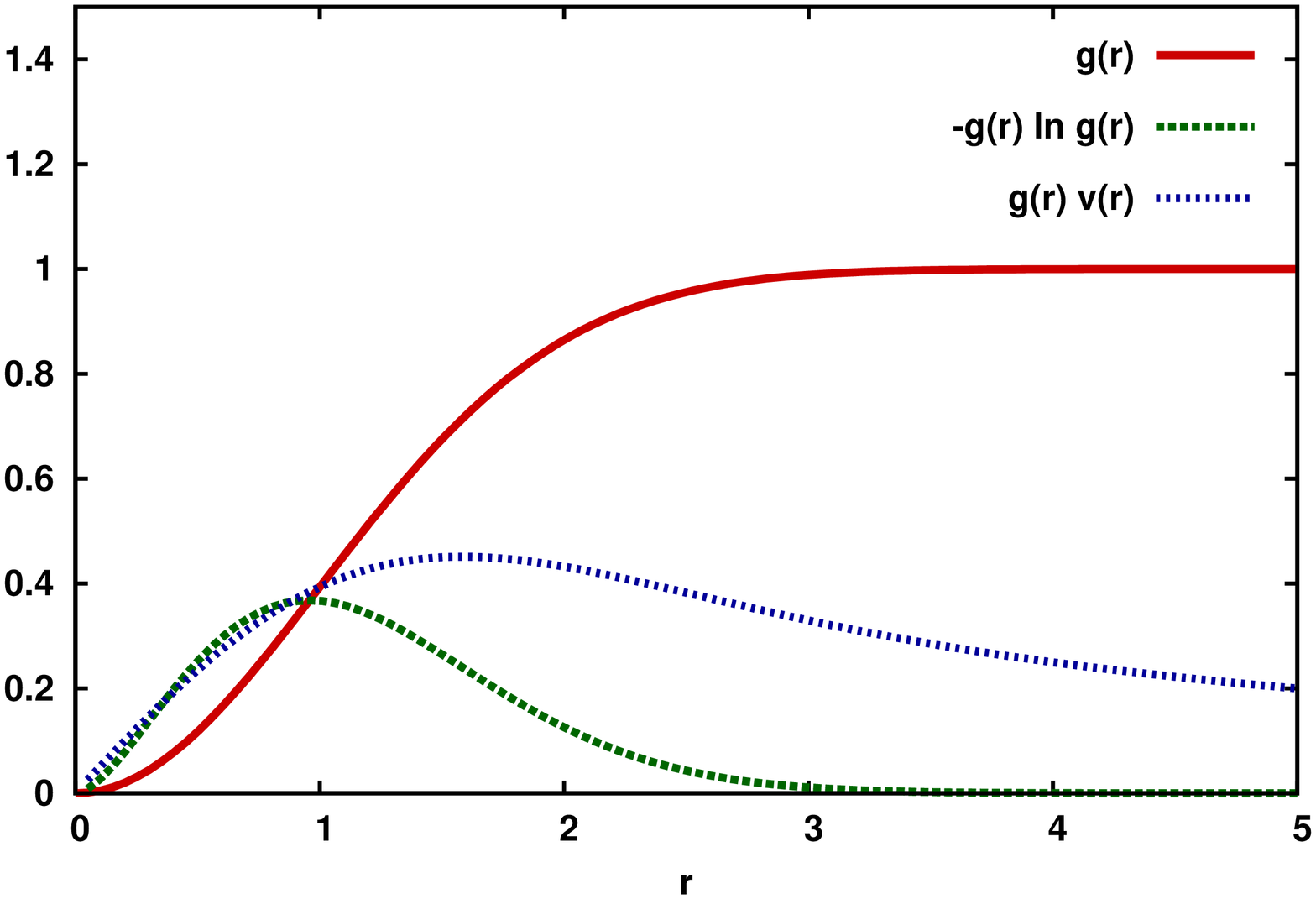}
    \includegraphics[width=0.25\textwidth,angle=-90]{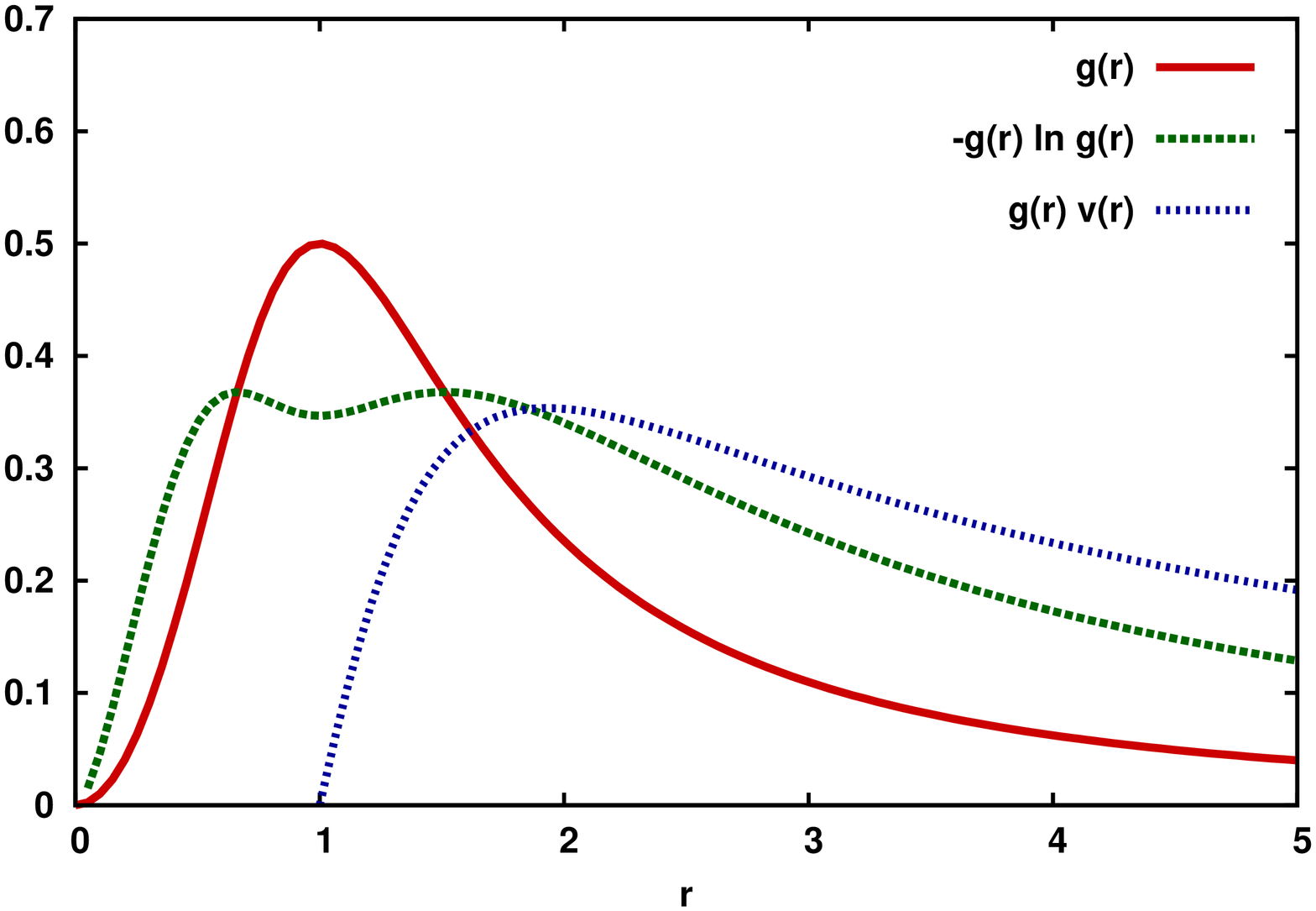}
  \end{center}
\caption{Schematic plots of the pair distribution function and the corresponding energy and entropy
contributions in short (top) and long range (bottom) correlated matter assuming a $1/r$ and a
 $\sigma r$ type pair potential, respectively.}
\label{fig:1}       
\end{figure}

The total energy can be calculated in a similar way. Assuming a $v(r_{12})$ pair-potential
depending on the relative coordinate only and individual kinetic energies, $K(p_i)$,
we arrive at
\be
 \frac{E_N}{N} \, = \,  \frac{\ebar}{\nbar} + \frac{\nbar}{2} V_{{\rm pair}}
\ee{SCALE-OF_ENERGY}
with
\ba
 \ebar &=& \pint{} f(p) K(p), \nl
 V_{{\rm pair}} &=& \rint{} g(r) v(r).
\ea{ENERGY-DENSITY}

Some typical $g(r)$ functions are shown in Fig.\ref{fig:1}. On the top figure a pair
distribution function tending to one at large distances and a Coulomb-like pair potential,
while on the bottom figure a linear confining potential, $v(r)  \sim r$
and a power-law tailed pair distribution function are assumed.

In most physical systems studied traditionally in thermodynamics, like gases, liquids, plasmas, etc.
the function $g(r)$ approaches the value one at large distances. Therefore there are no
non-extensive contributions to the entropy per particle. 
For the energy the situation is different so far, since $v(r)\sim r^{-b}$ might not approach zero
for large distances fast enough, producing this way a contribution to the energy per particle,
$E_N/N$ which may even diverge in the large $N$ limit. Such a case is an unscreened $1/r$-like
potential in three dimensions.

\section{Additivity, extensivity and abstract composition rules}
\label{sec:1}

Non-extensive quantities, whose amount per particle is not finite in the thermodynamic limit,
are also not additive, because the repeated composition by simple addition rules always
leads to a result proportional to the number of steps. It is possible, however, 
that one is able to find another quantity, a certain function of the non-additive one,
which is additive. This way the non-extensive thermodynamics can be treated by mathematical
algorithms which were designed for additive composition rules.
In this section we analyze the mathematical background of composing energy and/or entropy
of subsystems and then repeating this composition. The thermodynamical limit is
appointed to the infinite repetition of the composition with an infinitesimal amount \cite{EPLBiro}.

\subsection{General rules and thermodynamical limit}
\label{subsec:1.1}

Let us denote an abstract pairwise composition rule by the mapping 
$(x,y) \rightarrow h(x,y)$. 
The important question arises, that what happens if we repeat  such a
composition rule arbitrarily long, each time applying to an infinitesimal amount: 
This way one deals with the thermodynamical limit of composition rules corresponding to the energy
or the entropy. The effective rule in this limit, which applies to results of repeated
rules themselves, has special properties then.

From the starting rule we demand only a trivial property: that the composition with
zero should be the identity
\be
 h(x,0) = x.
\ee{TRIVIAL}
We do not assume in general symmetry (commutativity), such as $h(y,x)=h(x,y)$
nor we demand associativity
\be
 h(h(x,y),z) \: = \: h(x,h(y,z)).
\ee{ASSOC}
Her we note that the general solution of the associativity equation (\ref{ASSOC}) is given by
\be
 h(x,y) = X^{-1}\left( X(x)+X(y)\right)
\ee{FORMLOG}
with  $X(x)$ being a strict monotonic function \cite{SPANISH}. 
We shall refer to this mapping function
as the ''formal logarithm'',
because it maps the arbitrary composition rule $h(x,y)$ to the addition by taking the 
$X$-function of eq.(\ref{FORMLOG}):
\be
 X(h(x,y)) = X(x) + X(y).
\ee{FORMADD}
Due to this construction the generalized analogs to classical extensive (and additive) quantities
are their formal logarithms, whenever the composition rule is associative. As a consequence stationary
distributions, in particular those obtained by solving generalized Boltzmann equations \cite{NEBE}, 
are the Gibbs exponentials of the formal logarithm,
\be
 f(x) = \frac{1}{Z} e^{-\beta X(x)}.
\ee{GIBBSFORM}

Let us now regard a large number of iterations, $N$, of a general composition rule.
We  apply it to a small amount $y/N$ and repeat this $(N-1)$ times, constructing this way
the quantity
\be
 x_N(y) := \underbrace{h \circ\ldots \circ h}_{N-1} \left(\frac{y}{N}, \ldots, \frac{y}{N}\right). 
\ee{ITER}
We consider the large-$N$  limit, 
\be
 \lim_{N\rightarrow\infty}\limits x_N(y) < \infty,
\ee{LIM}
if this is finite for a finite $y$,
we  can apply all formulas of classical thermodynamics usually
applied to extensive quantities. 
Such a limiting quantity is  extensive, but not necessarily additive.
Our goal is to obtain the asymptotic composition rule,
\be
 x_{N_1+N_2} = \varphi(x_{N_1},x_{N_2})
\ee{ASYMPCOMP}
in the limit $N_1, N_2 \rightarrow \infty$.
The recursion for the $n$-th step of this repetitive composition is given by
\be
 x_n = h\left(x_{n-1},\frac{y}{N}\right),
\ee{RECURR}
starting with $x_0=0$. 
Subtracting $x_{n-1}=h(x_{n-1},0)$ from both sides we arrive at
\be
 {x_n-x_{n-1}} = {h(x_{n-1},\frac{y}{N})-h(x_{n-1},0)}.
\ee{DISCRETE}
Denoting by $t=(n-1)/N$ the extensivity share already achieved, one step
takes $\Delta t=1/N$, and the above recursion can be Taylor-expanded for
a small $y/N = y \Delta t$:
\be
 x(t+\Delta t) - x(t) = y \, \Delta t \, \left. \pd{}{y} h\left(x(t),y\right) \right|_{y=0^+}
 + {\cal O}(\Delta t ^2).
\ee{TAYLOR-EXPANDED-RECURSION}
In the large $N$ ($\Delta t \rightarrow 0$)
limit this becomes equivalent to a differential equation
similar to a renormalization flow equation:
\be
 \frac{dx}{dt} = {y} \: h_2'(x,0^+).
\ee{FLOW}
In this expression 
$h_2'(x,0^+)$ denotes the partial derivative of the rule $h(x,y)$ with respect to its
second argument taken when this  value approaches zero from above. 
Note that the uniformity of subdivisions to $y/N$ is
not necessary; all infinitesimal divisions summing up to $y$ by $t=1$ lead to the same
differential flow equation.

The solution of eq.(\ref{FLOW}), 
\be
 L(x) = \int_0^x\limits \frac{dz}{h_2'(z,0^+)} \, = \, y \, {t},
\ee{ASYLOG}
defines the additive mapping of $x$, i.e. the formal logarithm $L(x)$.
By the help of this the following asymptotic composition rule arises:
\be
 x_{12} := \varphi(x_1,x_2) = L^{-1} \left( L(x_1)+L(x_2) \right);
\ee{ASYCOM}
it is already associative and commutative.
Commutativity is trivial and associativity is also easily proved:
\ba
 \varphi(\varphi(x_1,x_2),x_3) \: = \: L^{-1}\left(L(\varphi(x_1,x_2))+L(x_3) \right)  \nl
 \qquad = L^{-1}\left(L(x_1)+L(x_2)+L(x_3) \right) \qquad \qquad \: \nl
 \qquad = L^{-1}\left(L(x_1)+L(\varphi(x_2,x_3)) \right) \qquad \qquad \quad  \nl
  = \quad \varphi(x_1,\varphi(x_2,x_3)) \, . \qquad \qquad \qquad \qquad \: \: \:
\ea{ASSOC_PROOF}

It is interesting to check 
that {\em all associative rules are mapped to themselves in the above limit}.
Given an associative composition rule, $h(x,y)$, it possesses a
formal logarithm, $X(x)$, which is additive:
\be
 X\left(h(x,y)\right) = X(x)+X(y).
\ee{STRATASSOC}
Now taking the derivative of this equality with
respect to the second argument we obtain 
\be
X'(h)\, \partial h/\partial y = X'(y) 
\ee{DERIVH}
which taken at $y=0$ becomes
\be
 h_2'(x,0^+) = \frac{X'(0)}{X'(h(x,0))}.
\ee{FIDUCIAL_DERIV2} 
Due to the property $h(x,0)=x$ (equivalently $X(0)=0$)
the formal logarithm of the asymptotic composition rule is given by
\be
 L(x) = \int_0^x\limits \frac{X'(z)}{X'(0)}dz = \frac{X(x)}{X'(0)};
\ee{FORMAL_ASYMP}
it is proportional to the formal logarithm of the starting rule. Therefore
the asymptotic rule is exactly the same as we begun with: $\varphi(x,y)=h(x,y)$.
The freedom in a factor of the formal logarithm is used to set $X'(0)=1$.
This way any associative composition rule describes a limiting rule of a class of
non-associative rules. 


\subsection{Deformed logarithms and deformed exponentials}
\label{subsec:1.2}

The stationary distribution eq.(\ref{GIBBSFORM}) in the large-$N$
limit contains the formal logarithm, $L(x)$. In fact the composed function,
$e_a=\exp\circ L$ is the one, which is frequently called a {\it 'deformed exponential'}
in the literature. Its inverse, $\ln_a=L^{-1}\circ\ln$ is then the corresponding
{\it 'deformed logarithm'}. These functions are inverse to each other.
Further properties of the traditional exponential and logarithm functions are, however,
not automatically inherited. In particular reciprocals and negatives follow
different rules as we are used to.

\vs
In the particular case of scaling formal logarithms,
\be
 L_a(x) = \frac{1}{a} L_1(ax),
\ee{SCALING_LOG}
several interesting identities hold, among others the followings:
\ba
 L_0(x) &=& x, \nl
 L^{-1}_a(x) &=& \frac{1}{a} L_1^{-1}(ax), \nl
 \ln_a(1/x) &=& - \ln_{-a}(x), \nl
 1/e_a(x) &=& e_{-a}(-x) 
\ea{INTERESTING}
Since $a=q-1$, the $a^*=-a$ duality corresponds to the $q^*=2-q$ Tsallis-duality.
This can be important for the particle-hole relation for fermions:
\be
 1 - \frac{1}{e_a(-x)+1} = \frac{1}{e_{-a}(x)+1}.
\ee{PTL_HOLE}

Let us now list some important particular rules and their asymptotic
pendants considered in applications of non-extensive statistics to physical systems.

The trivial (and classical) addition is the simplest composition rule: $h(x,y)=x+y$.
In this case $h_2'(x,0^+)=1$ and one obtains
\be
 L(x) = \int_0^x\limits dz = x.
\ee{ADDFORMALLOG}
The original Gibbs exponentials, $e^{-\beta E}/Z$, result as stationary
distributions from any Monte Carlo type algorithm using the additive composition rule. 
The asymptotic rule is also the addition $\varphi(x,y)=x+y$.

Another rule leading to the so-called 
$q$-exponential distribution \cite{GE5} is given by
$h(x,y)=x+y+axy$ with the parameter $a$ proportional to $q-1$ occurring
in the Pareto-Tsallis distribution. Now one obtains
$h_2'(x,0^+)=1+ax$ and
\be
 L(x) = \int_0^x\limits \frac{dz}{1+az} = \frac{1}{a} \ln(1+ax).
\ee{TSALLISLOG}
This formal logarithm leads to a stationary distribution with power-law tail
as the function composition $exp \circ L$ on the power $-\beta$:
\be
 f(E) = \frac{1}{Z} e^{- \frac{\beta}{a} \ln(1+a E)} = \frac{1}{Z} \left( 1+a E\right)^{-\beta/a}.
\ee{TSALLISEXP}
On the other hand, assuming such a non-additive composition rule for the generalized entropy, 
a special formula can be constructed
as the expectation value of the inverse of this function, of the deformed logarithm,
$L^{-1} \circ \ln$. One obtains
\be
 S = \int\! f \, \frac{e^{-a\ln(f)}-1}{a} \: = \: \frac{1}{a} \int  \, (f^{1-a}-f).
\ee{TSALLIS_ENTROPY}
The asymptotic composition rule again coincides with the original one:
$\varphi(x,y)=x+y+axy$. We note here that the formal logarithm of the integrated expression
is the (additive) R\'enyi entropy:
\be
 L(S) = \frac{1}{1-q} \ln\int\!f^q,
\ee{RENYI_ENTROPY}
with $a=1-q$ and $\int\!f=1$.

A further rule has been suggested by Kaniadakis\cite{KANIADAKIS}, 
based on the $\sinh$ function. The formal logarithm is given as
\be
 L(x) = \frac{1}{\kappa} {\rm Ar sh} (\kappa x),
\ee{KANI_LOG}
and its inverse becomes $L^{-1}(t)=\sinh(\kappa t)/\kappa$.
The stationary distribution, composed by \hbox{$exp \circ L$}, is
\be
 f_{{\rm eq}}(p) = \frac{1}{Z} \left(\kappa p + \sqrt{1+\kappa^2p^2}\right)^{-\beta/\kappa}.
\ee{KANI_DIST}
For large arguments it gives a power-law in the momentum $p$ and hence also in the
relativistic energy.
The corresponding entropy formula is the average of $L^{-1}\circ \ln$ over the allowed phase space:
\be
 S_K = - \int \frac{f}{\kappa} \sinh(\kappa\ln f) = \int \frac{f^{1-\kappa}-f^{1+\kappa}}{2\kappa}.
\ee{KANI_ENT}
The composition formula can be reduced to
\be
 h(x,y) = x\sqrt{1+\kappa^2y^2} + y\sqrt{1+\kappa^2x^2}.
\ee{KANI_ADD}
For low arguments it is additive, $h(x,y) \approx x+y$, for high ones it is multiplicative,
$h(x,y) \approx 2\kappa xy$. It has been motivated by the relativistic kinematics of
massive particles. 
Interpreting the parameter as $\kappa  = 1/mc$, one deals with $\kappa p = \sinh \eta$, 
so the formal logarithm becomes proportional to the rapidity, $L(p)=mc\eta$.
This implies a stationary distribution like $exp(-\beta mc\eta)$, which has not
yet ever been observed in particle spectra stemming from relativistic heavy ion collisions.
For such a purpose it is tempting to consider some further scenarios based on other quantities
than suggested above (see next section).

The rule leading to a stretched exponential stationary distribution, often considered in
problems related to anomalous diffusion and Levy-flights, is given by
$h(x,y)=\left(x^b+y^b\right)^{1/b}$. Here the partial derivative
is evaluated at a small positive argument, $\epsilon=y/2N$.
One obtains $h_2'(x,\epsilon)=c(\epsilon)x^{1-b}$ with a factor depending on $\epsilon$
and for given values of $b$ diverging in the $\epsilon=0$ limit. However, this
can be accommodated by our procedure; we obtain the formal logarithm $L(x)=c(\epsilon)x^b/b$, 
and therefore the asymptotic rule  $\varphi(x,y)=\left(x^b+y^b\right)^{1/b}$. Again, constant
factors in the formal logarithm can be eliminated without loss of any information.

Now let us investigate a non-associative rule; its asymptotic limit cannot be itself.
We regard a linear combination of arithmetic and harmonic means: 
\be
 h(x,y) = x + y + a \frac{xy}{x+y}
\ee{NONASSOC}
The rescaling flow derivative is given by
$h_2'(x,0^+)=1+a$ and -- being a constant -- it leads to $L(x)=x/(1+a)$ and
with that to the {\em addition} as the asymptotic rule: $\varphi(x,y)=x+y$.

As an interesting rule we discuss the relativistic formula for collinear velocity composition,
\be
 h(x,y) = \frac{x+y}{1+xy/c^2}.
\ee{EINSTEIN}
This rule is associative, and it also preserves its form in the thermodynamic limit.
The fiducial derivative is given by $h_2'(x,0^+)=1-x^2/c^2$ and the formal logarithm,
$L(x)= \, c {\rm \, atanh\: } (x/c)$ turns out to be the rapidity. The asymptotic composition
rule recovers the original one.

There are also general types of composition rules, which mutate into a simpler asymptotic form.
For our discussion particularly important are rules of the form
\be
 h(x,y) = x + y + G(xy)
\ee{IMPORTANT}
with a general function $G(z)$, restricted by the property $G(0)=0$ only.
In this case $h_2'(x,0)=1+G'(0)x$ asymptotically leads to a Tsallis-Pareto distribution
with the parameter $q-1=G'(0)$.

\subsection{Extreme relativistic kinematics}
\label{subsec:1.3}


In this section we review a particular type of pair interaction, which can be
expressed as a function of the kinetic energies of the individual particles.
The relation to relativistic kinematics is established by the fact, that
we consider such dependence through the Lorentz-invariant relative four-momentum
square variable:
\be
 E_{12} = E_1 + E_2 + U(Q^2).
\ee{RELATIVISTIC}
We study whether relativistic speeds
alone can cause ''non-extensivity'', i.e. a power-law tailed kinetic energy distribution. 
The relativistic formula for $Q^2$ is given by:
\be
 Q^2 = (\vec{p}_1-\vec{p}_2)^2 - (E_1-E_2)^2
\ee{Q2_LORENTZ}
with $\vec{p}_i,E_i$ being relativistic momenta and full energies of interacting bodies.
Expressed by the energies and the angle $\Theta$ between the two momenta this becomes
a linear expression of $\cos\Theta$:
\be
 Q^2 = 2 \left(E_1E_2 - p_1p_2\cos\Theta   \right) - (m_1^2+m_2^2)
\ee{Q2_THETA}
with $p_i=\sqrt{E_i^2-m_i^2}$ for $i=1,2$. Here we use relativistic units ($c=1$) and assume the 
masses $m_1$ and $m_2$, respectively, for the interacting partners. 
It is useful to note that writing eq.(\ref{Q2_THETA})
as $Q^2=2(A-B\cos\Theta)$ we have
\be
 A \pm B = E_1E_2-\frac{1}{2}(m_1^2+m_2^2) \pm p_1p_2. 
\ee{AB}

For the sake of simplification we average over the relative
directions of the respective momenta and obtain
\ba
 \langle U(Q^2) \rangle &=& \frac{1}{2} \int_{0}^{\pi}\limits U(2A-2B\cos\Theta) \sin\Theta \, d\Theta
\nl
& =& \frac{F(2A+2B)-F(2A-2B)}{4B}, 
\ea{U_ISO}
with $U(w)=dF/dw$. It is easy to derive by the substitution \hbox{$w=2(A-B\cos\Theta)$.}
The rule for the kinetic energy, $K_i=E_i-m_i$, composition is given by
\be
 K_{12} = K_1 + K_2 + \frac{F(2A+2B)-F(2A-2B)}{4B} 
\ee{KIN_COMPOSE}
The quantities $A$ and $B^2$ can be expressed by the respective kinetic energies and masses:
\ba
 A &=& K_1K_2+ (m_2K_1+m_1K_2) - \frac{1}{2}(m_1-m_2)^2, \nl
 B^2 &=& K_1K_2(K_1+2m_1)(K_2+2m_2). 
\ea{AB_KIN}
One observes that the product of kinetic energies occurs due to kinematic reasons.


Taylor expanding the integral of the unknown function $U(w)$ around $w=2A$
and ensuring the $h(x,0)=x$, as well as the $h(0,y)=y$ property,
we obtain the following composition rule for the relativistic kinetic energies:
\ba
 h(x,y) &=& x + y - U(2m_2x+m_{12}) - U(2m_1y+m_{12})  \nl
 && + U(m_{12}) +\sum_{j=0}^{\infty} U^{(2j)}(2A) \, \frac{(4B^2)^j}{(2j+1)!}
\ea{KINETIC_COMPOSITION}
with $m_{12}=-(m_1-m_2)^2$, 
\hbox{$A=xy+(m_2x+m_1y)+m_{12}/2$} and \hbox{$4B^2=4xy(x+2m_1)(y+2m_2)$.}
For unequal masses, $m_1 \ne m_2$ this composition rule is not symmetric.
Since
\ba
 \frac{\partial A}{\partial y}(x,0) = m_1+x, \nl
 \frac{\partial B^2}{\partial y}(x,0) = 2m_2x(x+2m_1),
\ea{AB_DERIVATIVES}
the  derivative leading to the formal logarithm of the asymptotic rule 
becomes an expression with a finite number of terms
\ba
 h_2'(x,0) &=& 1 - 2m_1 \, U'(m_{12}) + 2(m_1+x) \, U'(z) \nl
   && + \, \frac{4}{3} m_2 x \, (2m_1+x) \, U''(z),
\ea{FIDUC_KINETIC}
with $z=2A(x,0)=2m_2x+m_{12}$.
In all traditional approaches the interaction energy $U$ is independent
of $Q^2$. In such cases $h_2'(x,0)=1$ and the simple addition is the asymptotic composition
rule. Therefore the stationary energy distribution is of Boltzmann-Gibbs type.
For $Q^2$ dependent interactions on the other hand it is important to consider 
the extreme relativistic kinematics.
In this case the replacement $m_1=m_2=0$ leads directly to
\be
 h_2'(x,0) = 1 + 2x \, U'(0).
\ee{FIDUC_EXTREME}
As discussed in the previous subsection 
this generates a Tsallis-Pareto distribution in the relativistic kinetic energy.
This result includes for $U'=0$ the traditional momentum independent interaction
case leading to the addition as asymptotic
rule for non-relativistic kinetic energies, and hence to the Boltzmann-Gibbs distribution.
We note that in the relativistic kinematics the linear assumption, 
$U'=\alpha = {\rm const.}$  also leads to a 
Tsallis-Pareto distribution due to $h_2'(x,0)=1 + 2\alpha x$.

\subsection{Generalized entropies to each composition rule}
\label{sec:2.4}

There are two possible approaches in constructing a generalized entropy formula:
i) either to use a non-additive entropy for independent events with factorizing probability,
or ii) to search for an additive entropy while the common probability is not factorizing
in the individual probabilities. In both cases the entropy density function, $\sigma(p)$
to a probability $p$ can be obtained from the composition rule $h(x,y)$.

First we consider a non-additive entropy formula for factorizing probabilities, i.e.
\be
 \sum_{i,j} w_{ij} \sigma(w_{ij}) = h\left(\sum_i p_i\sigma(p_i), \sum_j q_j \sigma(q_j)\right)
\ee{FACTORNONADD}
with $w_{ij}=p_iq_j$. We would like to construct the function $\sigma(p)$ by knowing $h(x,y)$.
Let us inspect the equipartition case, $p_i=1/N_1$, $q_j=1/N_2$. In this case $w_{ij}=1/(N_1N_2)$.
Eq.(\ref{FACTORNONADD}) leads to
\be
 \sigma(ab) = h\left(\sigma(a),\sigma(b)\right)
\ee{SIGMAFUN}
with $a=1/N_1$ and $b=1/N_2$.
This requires the same composition rule for micro- and macro-entropy:
\be
 \sum_{ij} p_iq_j h(\sigma(p_i),\sigma(q_j)) =  h\left(\sum_i p_i\sigma(p_i), \sum_j q_j \sigma(q_j)\right).
\ee{MICROMACRO}
This $h$-extensivity can so far only be satisfied by the Tsallis rule $h(x,y)=x+y+axy$.
On the other hand if $q_0=1$ and all other $q_j=0$ for $j\ne 0$, we obtain two constraints:
\ba
 \sigma(p_i) &=& h(\sigma(p_i),\sigma(1)) \nl
 \sigma(0)   &=& h(\sigma(p_i),\sigma(0))
\ea{SIGMACONS}
from which it follows $h(x,0)=x$ with $\sigma(1)=0$ (the unexpectedness of a sure event is zero)
and $\sigma(0)=\infty$, too.

Based on the properties of the known $h(x,y)$, in the thermodynamical limit it is associative and
hence possess a formal logarithm, $L(x)$. Therefore
\be
 L(\sigma(ab)) = L(\sigma(a))+L(\sigma(b)),
\ee{AHA}
whose general solution is given by $L\circ\sigma=\beta\ln$. According to the tradition $\beta=-1$
in units of the Boltzmann constant, $k_B=1$, 
and therefore the entropy density function is expressed by the deformed logarithm:
\be
 \sigma(p) = L^{-1}(-\ln p) = \ln_a\left(\frac{1}{p}\right).
\ee{SIGMALOG}

It is possible to ask another question: if the construction rule for the common probability
is not the simple product, but it is known, what should the entropy density function be
in order to lead to the addition rule for the total entropy. So given the formula
\be
 w_{ij} = e^{h(\ln p_i, \ln q_j)}
\ee{NONFACT}
how to construct $\sigma(p)$ such that
\be
 \sum_{ij} w_{ij} \sigma(w_{ij}) = \sum_i p_i \sigma(p_i) + \sum_j q_j \sigma(q_j)
\ee{ADDISIG}
is fulfilled.
We have
\be
 c = w_{ij} = e^{h(\ln a, \ln b)}
\ee{COMPPROB}
as the known composite probability and seek for the entropy density function, $\sigma(a)$
satisfying
\be
 c \, \sigma(c) = ab \left( \sigma(a) + \sigma(b) \right).
\ee{SIFUNCTIONAL}
We solve this functional equation by deriving with respect to $b$ and take the result at $b=1$.
Since
\be
 \pd{c}{b} = e^{h(\ln a,\ln b)} \, h_2'(\ln a, \ln b) \frac{1}{b}
\ee{CDER}
we arrive at
\be
 a h_2'(\ln a, 0) \left( \sigma(a) + a \sigma'(a) \right) = a\sigma(a) + a \sigma(1) + a\sigma'(1).
\ee{FUDER}
Using  now that $\sigma(1)=0$ and $h_2'(x,0)=1/L'(x)$ with the formal logarithm $L$ associated to
the composition rule, $h$, we obtain -- using the variable $x=\ln a$ --
\be
 \frac{d\sigma}{dx} + \sigma = L'(x) \left( \sigma + \beta \right),
\ee{DFUDER}
with $\beta=\sigma'(1)$ constant. The final solution is expressed by the formal logarithm
of the asymptotic rule as
\be
 \sigma(a) = \beta e^{L(\ln a)-\ln a} \int_0^{\ln a} \! L'(u) \, e^{u-L(u)} \, du.
\ee{ADDITIVESIGMA}
It is interesting to note, that using $L^{-1}_a$ of the deformed logarithm
as the function $L$ belonging to the product composition rule
(\ref{COMPPROB}), one assumes \hbox{$\ln_a(w_{ij})=\ln_a(p_i)+\ln_a(q_j)$} and arrives at
\be
 p \: \sigma(p) = - e^{\ln_ap} \, \int_0^{\ln_ap} e^{-u} e_a(u) du.
\ee{DEFENTROPY}

\section{Non-extensive Boltzmann equation}
\label{sec:2}


\vs

\vs
It is a false belief that only the exponential distribution can be the stationary
solution to the Boltzmann equation: this statement is true only  
i) if the two-particle distributions factorize, ii) the two-particle energies are additively
composed from the single-particle energies ($E_{12}=h(E_1,E_2)=E_1+E_2$ ) and 
iii) the collision rate is multilinear in the two-particle (and two-hole) densities.
A generalization of the original Boltzmann equation has been pioneered by
Kaniadakis\cite{BOLTZMANN-GENERAL} investigating  nonlinear
density dependence of the collision rates. An $''H_q''$ theorem for the particular
Tsallis form of the collision rate has been derived by Lima, Silva and 
Plastino\cite{HQ-THEOREM}.

\vs
A possible generalization of the Boltzmann equation uses an altered form of
the 'Stosszahlansatz' and allows for an evolution equation of a function of the
original phase space occupation factor, $F(f)$:
\be
 D F(f_1) = \int_{234}\limits w_{1234} \left( G_{34} - G_{12}\right)
\ee{GENER_BOLTZMANN}
with
\be
 D F =  \frac{p^{\mu}}{p^0} \partial_{\mu} F
\ee{DERIV}
total (Vlasov-) derivative, with a $1234$-symmetric collision rate including Dirac-delta
distributions for momentum and energy composition rules in two-to-two collisions (which also may be
of generalized type by using corresponding formal logarithms), and finally the
generalized product for the two-particle density factor,
\be
 G_{12} = e_a \left( \ln_a(f_1) + \ln_a(f_2) \right)
\ee{TWOPART}
using the deformed exponential and logarithm functions.
Based on this, a particular expression for the entropy current density can be
defined:
\be
 S^{\mu} = - \int \frac{p^{\mu}}{p^0} \, \sigma(F(f)).
\ee{H_ENTROPY}
The entropy density form, $\sigma(F)$ always can be constructed in a way, that the
second theorem of thermodynamics is fulfilled.
The local source for the entropy is namely given by
\be
 \partial_{\mu} S^{\mu} = - \int_1\limits \sigma'(F(f_1)) \, DF(f_1).
\ee{ENTROPY_SOURCE}
Utilizing the generalized Boltzmann equation (\ref{GENER_BOLTZMANN}) and exchanging the
index $1$ with $2$, $3$ and $4$ while $w_{1234}$ stays invariant and obviously
$G_{ij}=G_{ji}$, one arrives at
\be
 \partial_{\mu} S^{\mu} = \frac{1}{4} \int_{1234}\limits  w_{1234}
 \left( \sigma'_1 +\sigma'_2 - \sigma'_3 - \sigma'_4 \right)
 \left( G_{12}-G_{34}\right)
\ee{SCRAMBLED}
with $\sigma'_i=\sigma'(F(f_i))$ for $i=1,2,3,4$.
This quantity is always non-negative, i.e. 
\be
 \left(\Phi(G_{12})-\Phi(G_{34})\right) \, (G_{12}-G_{34}) \, \ge \, 0,
\ee{GREATER_THAN_ZERO}
if and only if
\be
 \Phi(G_{12}) = \sigma'(F(f_1))+\sigma'(F(f_2))
\ee{IRREV}
is a monotonic rising function.
Inspecting the generalized Stosszahlansatz eq.(\ref{TWOPART}) one finds that this splitting
to the sum of respective functions of $f_1$ and $f_2$ is only possible, if
$\Phi(t)\propto \ln_a(t)$. Therefore we conclude that
\be
 \sigma'(F(f)) = \alpha \ln_a(f) + \beta
\ee{SMALL_SIGMA_PRIME}
with $\alpha \ge 0$ and $\beta$ undetermined constants. (This derivation followed the spirit of
Ref.\cite{BOLTZMANN-GENERAL}.)

\vs
The generalized entropy density as a function of the one-particle phase space occupation
density is hence given by
\be
 \sigma(f) =  \int  F'(f) \left( \alpha \ln_a(f) + \beta  \right) df .
\ee{SMALL_SIGMA}
The traditional Boltzmann formula arises for $F(f)=f$ and $\ln_a(f)=\ln(f)$ (i.e. $a=0$).
Lavagno et.al. \cite{Lavagnoetal} considered $F(f)=f^q$ and $\ln_a(f)=(f^{q-1}-1)/(q-1)$ (i.e. $a=(q-1)$
and Tsallis composition rule for $\ln f$). In the case of $h(x,y)=x+y+axy$
one considers $\ln_a(x)=(x^a-1)/a$, $e_a(t)=(1+at)^{1/a}$ and
$G_{12}=(f_1^a+f_2^a-1)^{1/a}$. For a small $a$ parameter it is 
$G_{12}\approx f_1f_2(1 - a \ln(f_1)\ln(f_2)+\ldots)$.

\vs
We note that the detailed balance distribution is given by the condition 
$G_{12}=G_{34}$, while the corresponding energy composition rule
applies $L(E_1)+L(E_2)=L(E_3)+L(E_4)$. This is possible only if
$\ln_a(f_i)=-(L(E_i)-\mu)/T$, so
\be
 f^{{\rm eq}}(E) = e_a\left(\frac{\mu-L(E)}{T} \right).
\ee{GENER_EQUIL}
The parameters $T$ and $\mu$ are arbitrary constants for being a stationary solution
of the generalized Boltzmann equation, but they can be related to the total energy and
particle number in a given application.

\vs
\subsection{Deformed energy composition rules in parton cascade}

Our fist numerical approach \cite{NEBE} was restricted to the use
of abstract composition rules in the energy balance part: 
we equated the energy of the reacting parts before and
after the collision via an abstract energy composition rule
\be
 h(E_1,E_2) \: = \: h(E_3,E_4).
\ee{THE-RULE}
Although this rule cannot be specified without further knowledge, according
to our results presented in the previous section, in the thermodynamical limit
an asymptotic rule can be considered, with a formal logarithm.
The parton cascade simulation based on a Boltzmann equation is hence
modified by considering
\be
 L(E_1) \, + \, L(E_2) \: = \: L(E_3) \, + \, L(E_4).
\ee{ADD-RULE}
At the same time we applied $F(f)=f$ and $a=0$.
Applying such a general energy composition rule considered in the thermodynamical limit,
the rate of change of the one-particle distribution is given by
\be
\dot{f}_1 = \int_{234}\!\!\!\!\!\! w_{1234} \,
 \left[ f_3f_4 - f_1f_2 \right].
\ee{BOLTZMANN-2-BODY}
with the symmetric transition probability $w_{1234}$ including the constraint
\be
\Delta \: = \: \delta^3(\vec{p}_1+\vec{p}_2-\vec{p}_3-\vec{p}_4) \,
	\delta\left(h(E_1,E_2) - h(E_3,E_4)\right).
\ee{GEN-CONSTR}

\begin{figure}
\begin{center}
\includegraphics[width=0.3\textwidth,angle=-90]{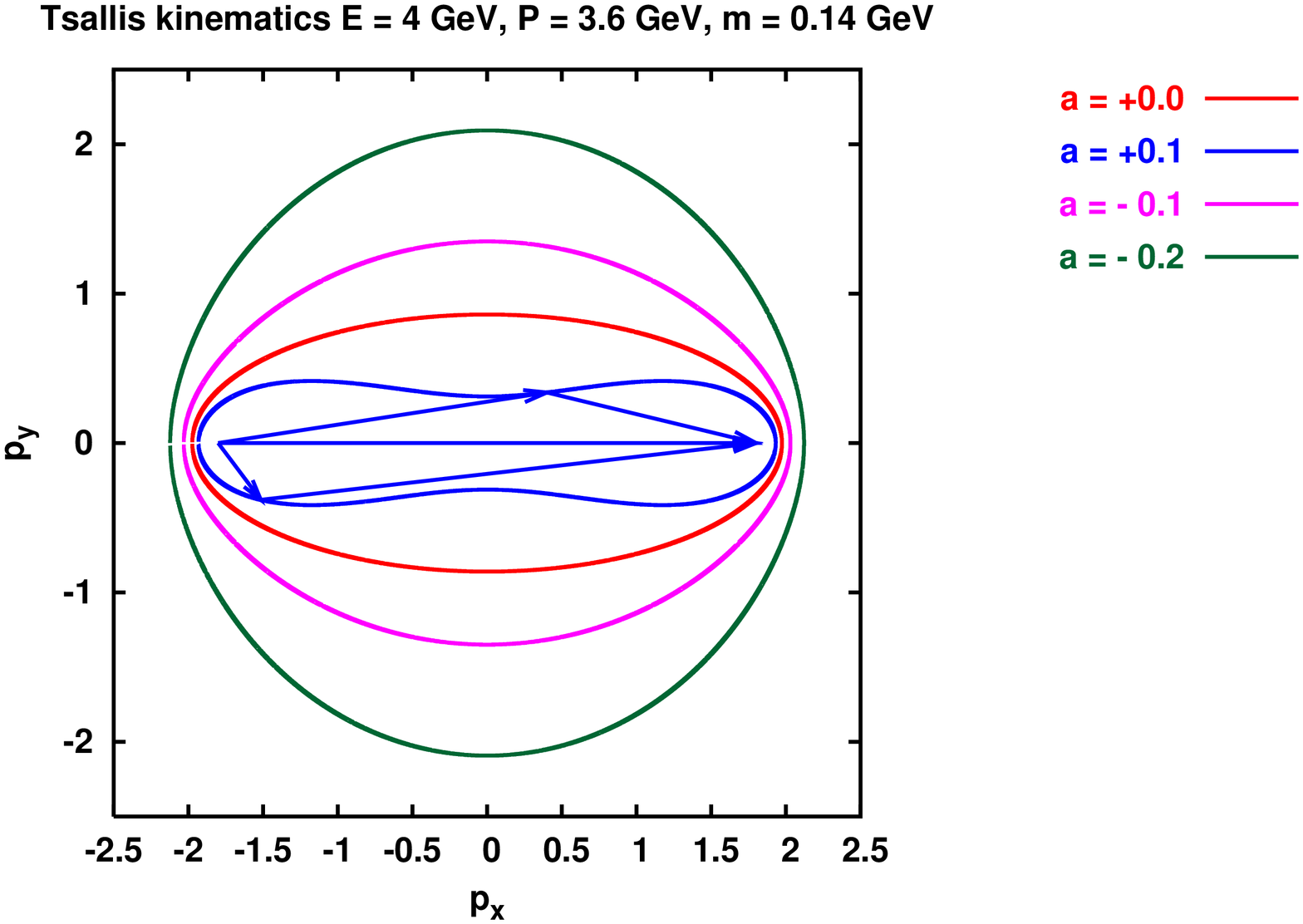}
\includegraphics[width=0.3\textwidth,angle=-90]{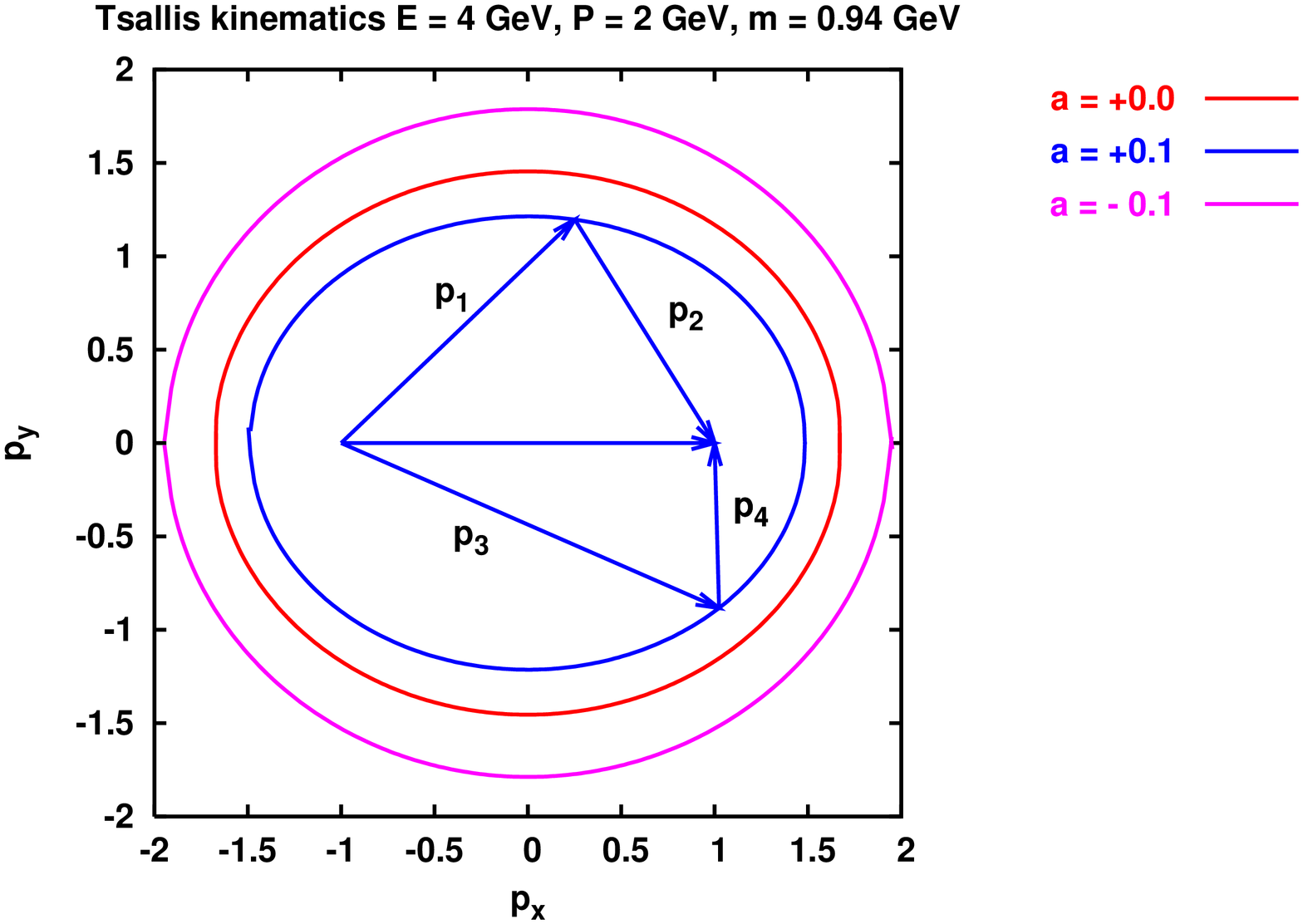}
\end{center}
\caption{ \label{Kinematic}
 Momentum vectors for pions (upper) and protons (lower) at $E=4$ GeV
 pair energy according to eq.(\ref{GEN-CONSTR}) with the rule
 $h(x,y)=x+y+axy$ for the energies.
}
\end{figure}

In the figure \ref{Kinematic} the possible pairs of momentum vectors are shown
for the $h(x,y)=x+y+axy$ energy composition rule for pions ($m=0.14$ GeV mass)
and protons ($m=0.94$ GeV mass), respectively. The two-dimensional cuts for
the endpoints of the respective vectors form an ellipsoid in the traditional
$a=0$ case, while this surface is deformed for nonzero extensivity parameters,
as seen in the figure.

In a stationary state the $f(E_i)$ distributions depend on the phase space points
through the energy variables only (this is to be checked on experimentally
observed hadron transverse momentum spectra at mid-rapidity by the so called $m_T$-scaling)  
and the detailed balance principle requires
\be
 f(E_1) \, f(E_2) \: = \: f(E_3) \, f(E_4).
\ee{EQUIL}
With the generalized constraint (\ref{GEN-CONSTR}) this relation is satisfied by
\be
 f(E) \: = \: f(0) \exp(-L(E)/T).
\ee{EQ-SOLUTION}
For the Tsallis-type energy addition rule\cite{TSALLIS-RULES,TSALLIS-WANG}, 
one obtains cut power-law stationary distribution,
\be
 f(E) \: = \: f(0) \left( 1 + bE \right)^{-1/bT}.
\ee{CUTPOWERLAW}
Connecting this to the Tsallis parametrization one uses $q=1+bT$.
Since the energy addition rule conserves in a two by two collision the quantity
$h(E_1,E_2)$, the new energies after such an event lie
on the $h(E_3,E_4)$=constant surface. Due to the  additivity of the formal logarithm of the
single particle kinetic energies,
$L(E_i)$, the total sum $L_{tot}=\sum_i L(E_i)$, is a conserved quantity.
This rule was applied in numerical simulations \cite{NEBE,PLAnoneq}. 
During the numerical searches for stationary distributions only the tacit assumption of constant
transition probability rates has been applied; the evolution results are obtained
in terms of the number of pairwise momentum exchange events, not in terms of real time.

\subsection{Random momenta}
\label{2.2a}

Parton cascade simulations usually consider pairwise collisions with energy and momentum
conservation inside the two-particle system. The pairs to collide are chosen randomly 
from an ensemble of particles and the new momenta are generated randomly according
to the above constraints. This way the probability is uniform in the two-particle phase space, 
provided the conditions for momentum and energy sums (in our more general case for the
energy composition) are satisfied:
\be
 d^2w = w_0 \delta(\vec{p}_1+\vec{p}_2-\vec{P}) \delta(h(E_1,E_2)-H) d^3p_1 d^3p_2.
\ee{PROBAB}
The constant $w_0$ is fixed by the normalization of the integral of this probability 
density to one (or to the actual collision rate in real-time simulations). 
Since there are six degrees of freedom and four constraints, two free
quantities have to be chosen randomly. It is, however, a delicate procedure to ensure
the random uniformity in the two-particle phase space for a general energy composition
rule.

It is customary to introduce the sum and difference of the momentum vectors by
\be
 \vec{p}_{1,2} = \frac{1}{2} \vec{P} \pm \vec{q}.
\ee{SPLITMOM}
Using this notation the momentum sum constraint can be integrated out trivially and
- since the Jacobean of the transformation (\ref{SPLITMOM}) is one - we arrive at
\be
 d^2w = w_0 \delta(h(E_1,E_2)-H) d^3q.
\ee{SIMPROB}
For the addition rule, $h(E_1,E_2)=E_1+E_2$, 
it is enough to obtain the direction of the vector $\vec{q}$ accordingly
while its magnitude is constrained by the energy sum. It is a straightforward task to do it
in the center of mass system, where the momentum sum vector, $\vec{P}$, vanishes:
the direction of the difference vector $\vec{q}$ in this system is uniform on a spherical surface.
A Lorentz-transformation into this system, a random azimuthal angle and a random 
cosine, and finally a back transformation provide the new momenta after a collision.

Since we are dealing with a constraint more general in the energy variables, first
we transform the problem of randomly choosing the difference vector $\vec{q}$ into a
problem of choosing proper energies after the collision. The energies are expressed
by the free dispersion relations
\be
 E_{1,2}^2-m_{1,2}^2 = \frac{1}{4} P^2 + q^2 \pm Pq\cos \theta,
\ee{MASS_SHELL}
where $P$ and $q$ denote the lengths of the corresponding vectors and $\theta$ the angle
between them. From this two equations one easily derives the following energy differentials:
\ba
 2E_1dE_1 &=& 2qdq + P\cos\theta dq - Pq \sin \theta d\theta, \nl
 2E_2dE_2 &=& 2qdq - P\cos\theta dq + Pq \sin \theta d\theta.
\ea{DIFFENER}
The phase space volume element can be expressed easily by using the wedge product form:
\be
 d^3q = dq \wedge q\sin\theta d\theta \wedge q d\phi 
\ee{WEDGE}
which upon using eq.(\ref{DIFFENER}) can be written as
\be
 d^3q = \frac{E_1E_2}{P} dE_1 \wedge dE_2 \wedge d\phi .
\ee{EWEDGE}
Now using the energy composition constraint we arrive at a probability density which is
not uniform in the energy:
\be
 d^2w = w_0 \delta(h(E_1,E_2)-H) \frac{E_1E_2}{P} dE_1 dE_2 d\phi.
\ee{PROBENER}
One uses the constraint to eliminate say $E_2$ from the above formula and considers
\be
 d^2w = w_0 \frac{E_1E_2}{Ph_2'(E_1,E_2)} dE_1 d\phi. 
\ee{AHAENER}
In the general case the differential probability density, $dw/dE_1$, is a complicated
function of the energy. Its integral, $w(E_1)$ has to be uniformly distributed.

In the case of a Tsallis composition rule one obtains \hbox{$E_2=(H-E_1)/(1+aE_1)$} 
and we arrive at
\be
 d^2w = \frac{w_0}{P} \frac{E_1(H-E_1)}{(1+aE_1)^2} \, dE_1 d\phi.
\ee{TSALLAHA}
This expression can be integrated giving
\be
 d^2w = \frac{1}{2\pi} d\rho \, d\phi,
\ee{INTEGTSALLAHA}
with
\be
 \rho(E_1,a) = \frac{(2+aH)\ln(1+aE_1)-\frac{aE_1}{1+aE_1}\left(2+aH+aE_1\right)}{(2+aH)\ln(1+aH)-2aH}
\ee{WHATISRHO}
when properly normalized.
The only problem is that $\rho(E_1)$ cannot be inverted
analytically. Even in the traditional case with $a=0$, the inversion requires
the solution of a third order equation:
\be
 \rho(E,a=0) = 3 (E/H)^2 - 2 (E/H)^3
\ee{RHOFORAZERO}
is distributed uniformly between zero and one. It means that  $E$ is between zero and $H$, the total
composed energy.

\begin{figure}
\begin{center}
\includegraphics[width=0.3\textwidth,angle=-90]{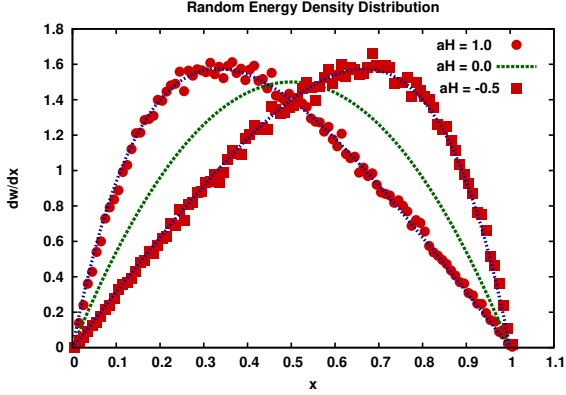}
\end{center}
\caption{\label{RED}
  The scaled differential probability density, $dw/dx$ is shown as a function of
  the random energy share of one of the collided particles $x=E_1/H$ (cf. eq.(\ref{TSALLAHA}))
  for different values of $aH$. Full circles and boxes denote the random energy deviates
  obtained numerically using the rejection method.
}

\end{figure}

After having $E_1$ and $E_2$ the momenta can be reconstructed with the help of a vector
triad describing the direction of the momentum sum, $\vec{P}$. They are given as
\hbox{$n_i = P_i / P$} and $\: e_i \: = \: (-n_{\perp}, \, n_1n_2/n_{\perp}, \, n_1n_3/n_{\perp})$ where 
the notation
\hbox{$n_{\perp}=\sqrt{n_2^2+n_3^2}$} stands for the component perpendicular to the first axis. 
The third orthogonal unit vector is $f_i=(0,-n_3/n_{\perp},n_2/n_{\perp})$. 
The momentum difference vector is hence reconstructed as
\be
 \vec{q} = q_{||} \: \vec{n} + q_{\perp} (\cos\phi \: \vec{e} + \sin\phi \: \vec{f} )
\ee{RECONSTRUCTEDq}
with
\ba
 q_{||} &=& \frac{E_1^2-E_2^2}{2P}, \nl
 q^2 &=& \frac{E_1^2+E_2^2}{2} - \frac{P^2}{4}, \nl
 q_{\perp} &=& \sqrt{q^2-q_{||}^2}.
\ea{qVECTOR}

\subsection{Parton cascade simulation}
\label{subsec:2.2}

\begin{figure}
  \begin{center}
	\includegraphics[width=0.25\textwidth]{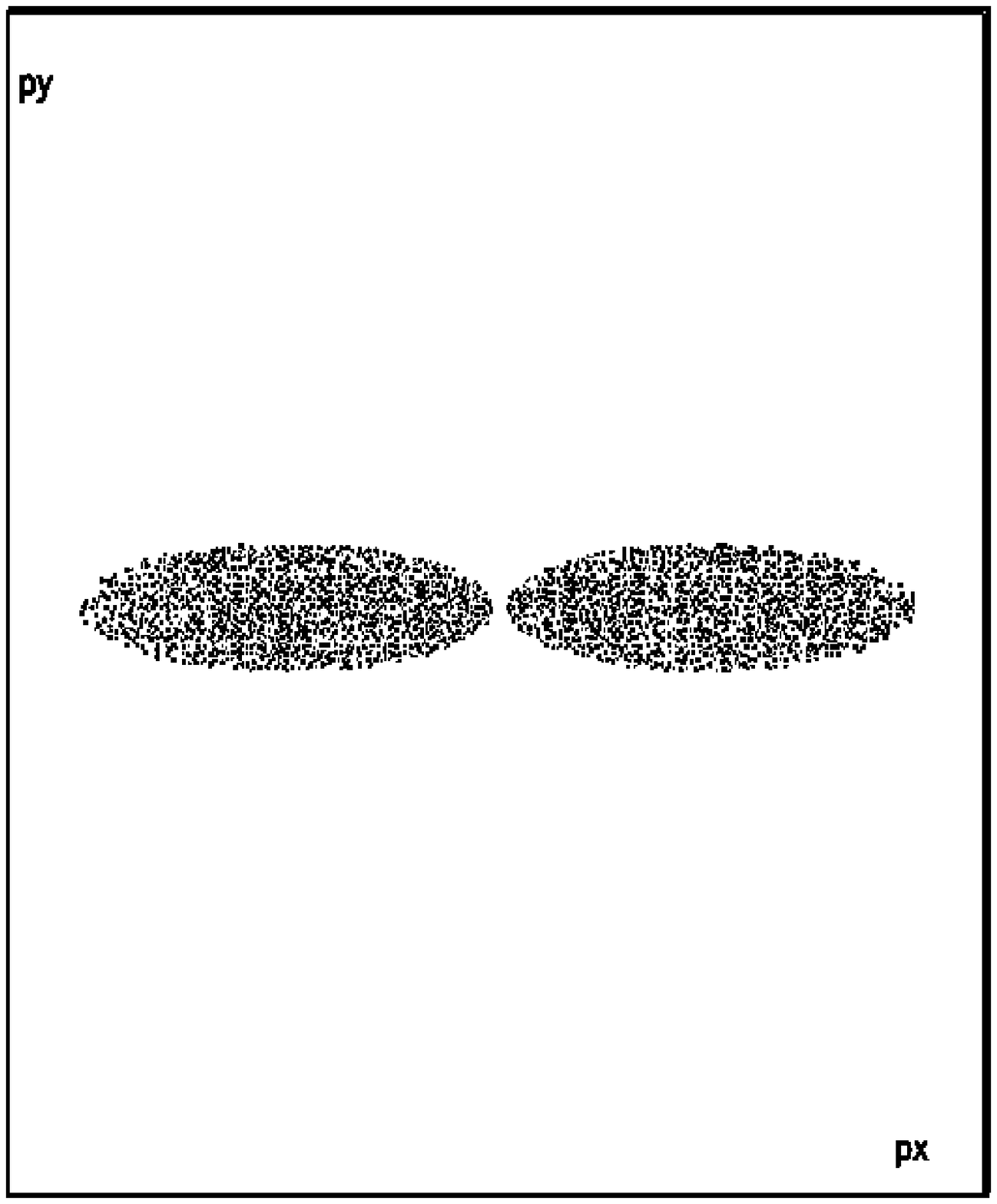}
	\includegraphics[width=0.25\textwidth]{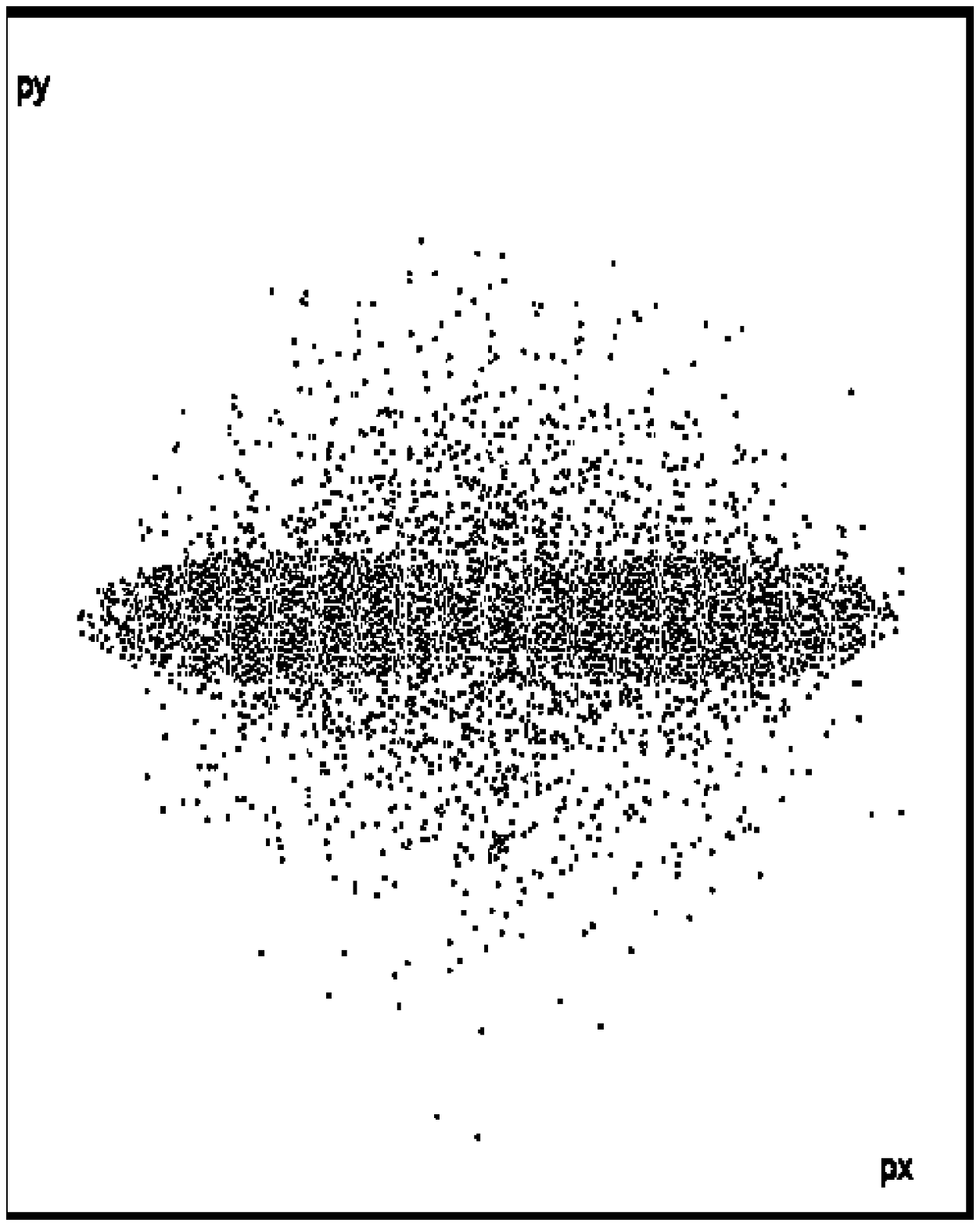}
	\includegraphics[width=0.25\textwidth]{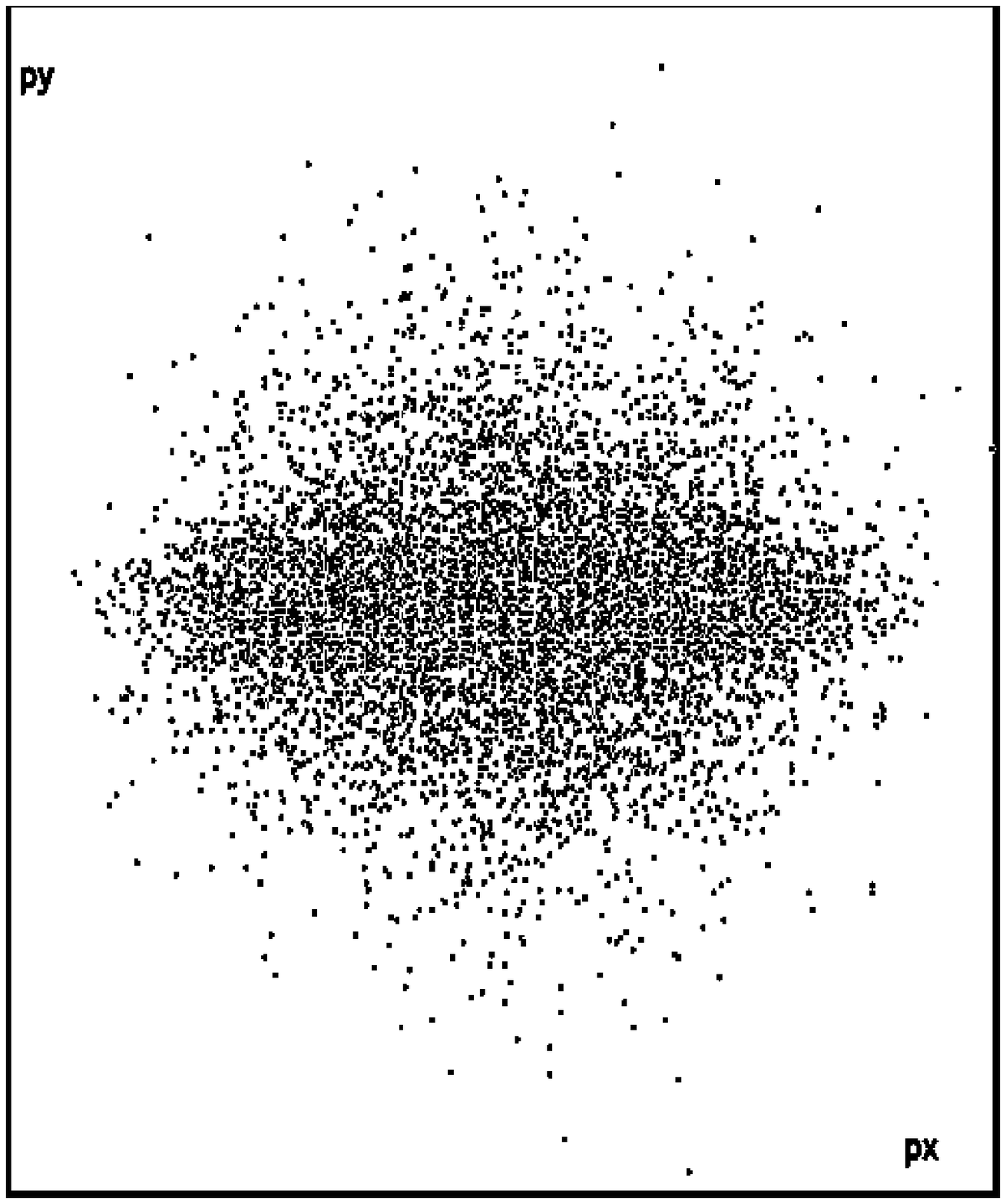}
	\includegraphics[width=0.25\textwidth]{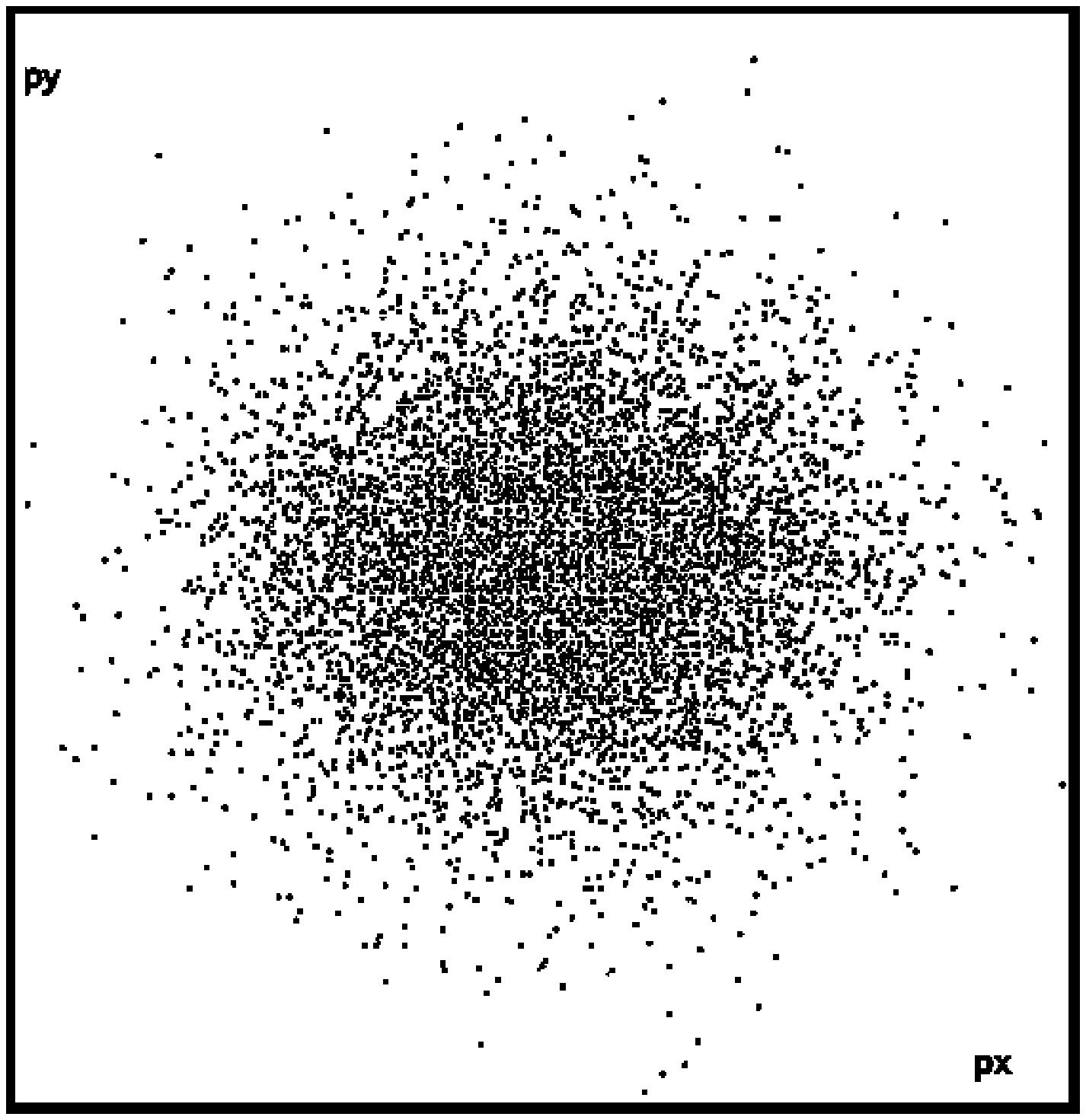}
  \end{center}
\caption{ \label{SNAPSHOTS}
 Snapshots of phase space cuts in the $p_x-p_y$ plane for colliding partons
 with the deformed energy composition rule $h(x,y)=x+y - 0.2 \: xy$ at $t=0, 0.3, 1,$ and $3$
 (from top to bottom). 
}
\end{figure}

First we show some snapshots of the colliding partons in the $p_x-p_y$ phase space
cut at different stages of the evolution marked by the average number of collisions
per particle, $t$ (cf. Fig.\ref{SNAPSHOTS}). 
At the beginning $t=0$ we prepared two distributions at a given
energy per particle and then Lorentz boosted each with $y_B=2$ units of rapidity in opposite
ways in the $p_x$-direction. The dark dots 
represent particle momenta stemming from the respectively boosted original sets.
The evolution towards a zero centered and isotropic distribution of momenta signals
already that thermal equilibration happens.

\begin{figure}
\begin{center}
\includegraphics[width=0.32\textwidth,angle=-90]{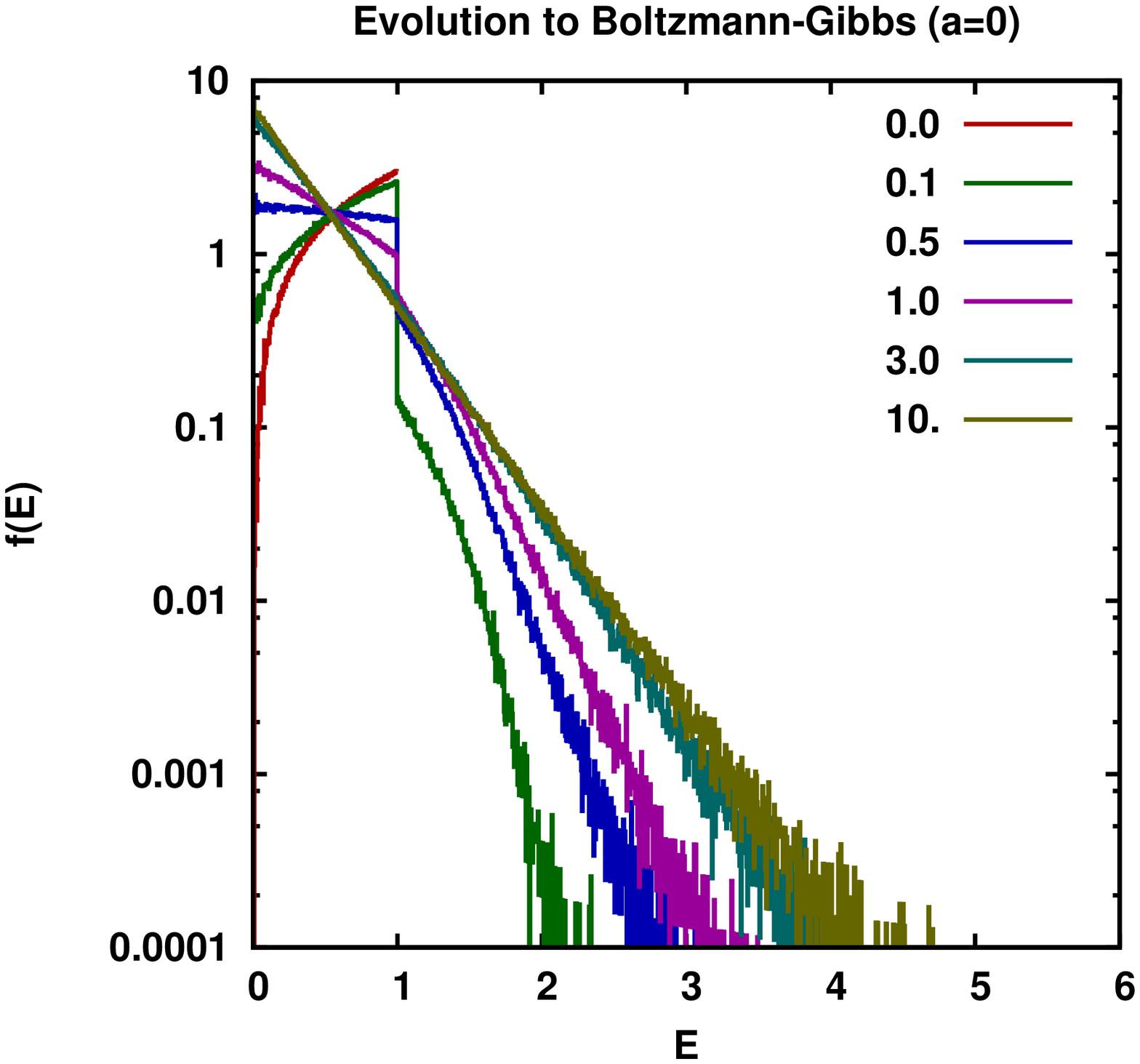}
\includegraphics[width=0.32\textwidth,angle=-90]{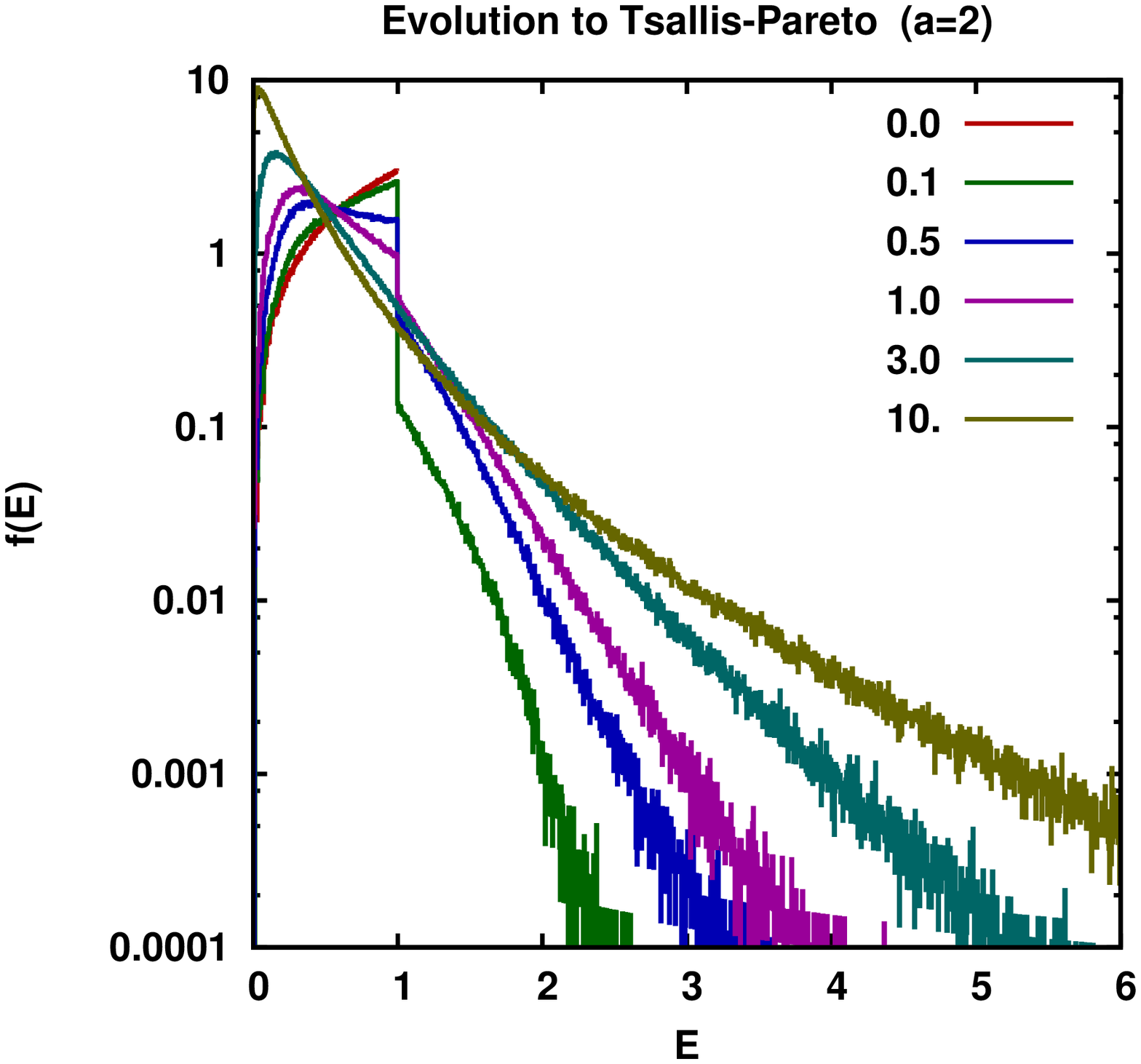}
 \end{center}
 \caption{
 \label{FIG-TS}
 	Evolution of single particle energy kinetic energy distributions for
        massless particles towards the Boltzmann-Gibbs distribution 
	for $h(E_1,E_2)=E_1+E_2$ (left part) and towards the
	Tsallis-Pareto one (right part) for $h(E_1,E_2)=E_1+E_2+2E_1E_2$.
	The curves are normalized to the same integral $\int E^2 f(E)dE$.
 }
 \end{figure}

\vs
Fig.\ref{FIG-TS} presents results of a simple test 
particle simulation with the  rule $h(x,y)=x+y+axy$ with $a=0$ (left) and $a=2$ (right), respectively.
We started with a uniform energy-shell distribution between
zero and $E_0=1$ with a fixed number of particles $N=10^6$.
The one-particle energy distribution evolves towards the
well-known exponential curve for $a=0$,
shown in the left part of Fig.\ref{FIG-TS}. These snapshots
were taken initially and after $0.1, 0.5, 1, 3$ and $10$ two-body collisions per particle.
Using the prescription with $a=2$, the energy distribution
approaches a Tsallis-Pareto distribution.

\vs

\vs
It is in order to make some remark on the energy conservation. 
For $h(x,y)=x+y$ we simulate a closed system with elastic collisions:
The sum, $E_{{\rm tot}}=\sum_{i=1}^N E_i$, does not change in any of the binary
collisions.  This is different by using a non-extensive formula 
for $h(x,y)$. 
With a constant positive (negative) $a$,
the bare energy sum is decreasing (increasing) 
while approaching the stationary distribution, while the sum of the
formal logarithms of the energy remains constant.
Open systems may gain or loose energy during their
evolution towards a stationary state.

\section{Non-extensive thermal equilibration}
\label{sec:3}



In order to investigate the equilibration of non-extensive systems we start with two subsystems, 
equilibrated separately. In order to prepare these systems the non-extensive Boltzmann equation 
can be solved numerically in a parton cascade simulation as described in the previous section. 
As an alternative way we use initial momentum distributions prepared by Monte Carlo rejection techniques
in the form of eq.(\ref{EQ-SOLUTION}), with different energy per particle but a common
parameter $a$ for the one an the other half of the particles.
Then random binary collisions between randomly chosen 
pairs of particles are evaluated. By doing so we apply the rules
\begin{equation}
	X\left(E_1\right) + X\left(E_2\right) = X\left(E_3\right) + X\left(E_4\right),  \label{bcr1}
\end{equation}
\begin{equation}
	\vec p_1 + \vec p_2 = \vec p_3 +\vec p_4.  \label{bcr2}
\end{equation}
In each step of the simulation we select two particles to collide. Then we find the value for 
the new momentum of the first particle (\(\vec p_3\)) satisfying the above constraints but 
otherwise random. Then applying eq.\ \ref{bcr2} we calculate the momentum of the second 
outgoing particle (\(\vec p_4\)). 
In these particular simulations we use the free dispersion 
relation for massless particles (\(E_i(\vec p_i) = \left|\vec p_i\right|\)),
since we are interested in the extreme relativistic kinematics case.
A typical simulation includes \(10^6-10^7\) collisions among \(10^5-10^6\) particles. 
After \(3-5\) collisions per particle on the average, the one-particle distribution 
approaches its stationary form sufficiently.

The following quantities are conserved during the simulation:
\begin{equation}
	X\left(E_{tot}\right) = \sum_{i=1}^N X\left(E_i\right), \;\;\;\;
	\vec P = \sum_{i=1}^N \vec p_i, \;\;\;\;
	N = \sum_{i=1}^N 1.
\end{equation}
We use the rule $h(x,y)=x+y+axy$ for the energy composition, here $a\sim (q-1)/T$  is  
the non-extensivity parameter. Our model reconstructs the traditional Boltzmann-Gibbs 
thermodynamics in the limit of \(a=0\).

As a preparation for the study of non-extensive thermal equilibration,
we perform simulations on two large  subsystems with particle numbers \(N_1=N/2\) and \(N_2=N/2\), 
total (quasi-)energies \(X(E_1)\) and \(X(E_2)\) and non-extensivity parameter $a$. 
The unified system is taken as an initial state with \(N=N_1+N_2\) particles.

\subsection{Simulation results}

\begin{figure}
\begin{center}
 	\includegraphics[width=0.32\textwidth,angle=-90]{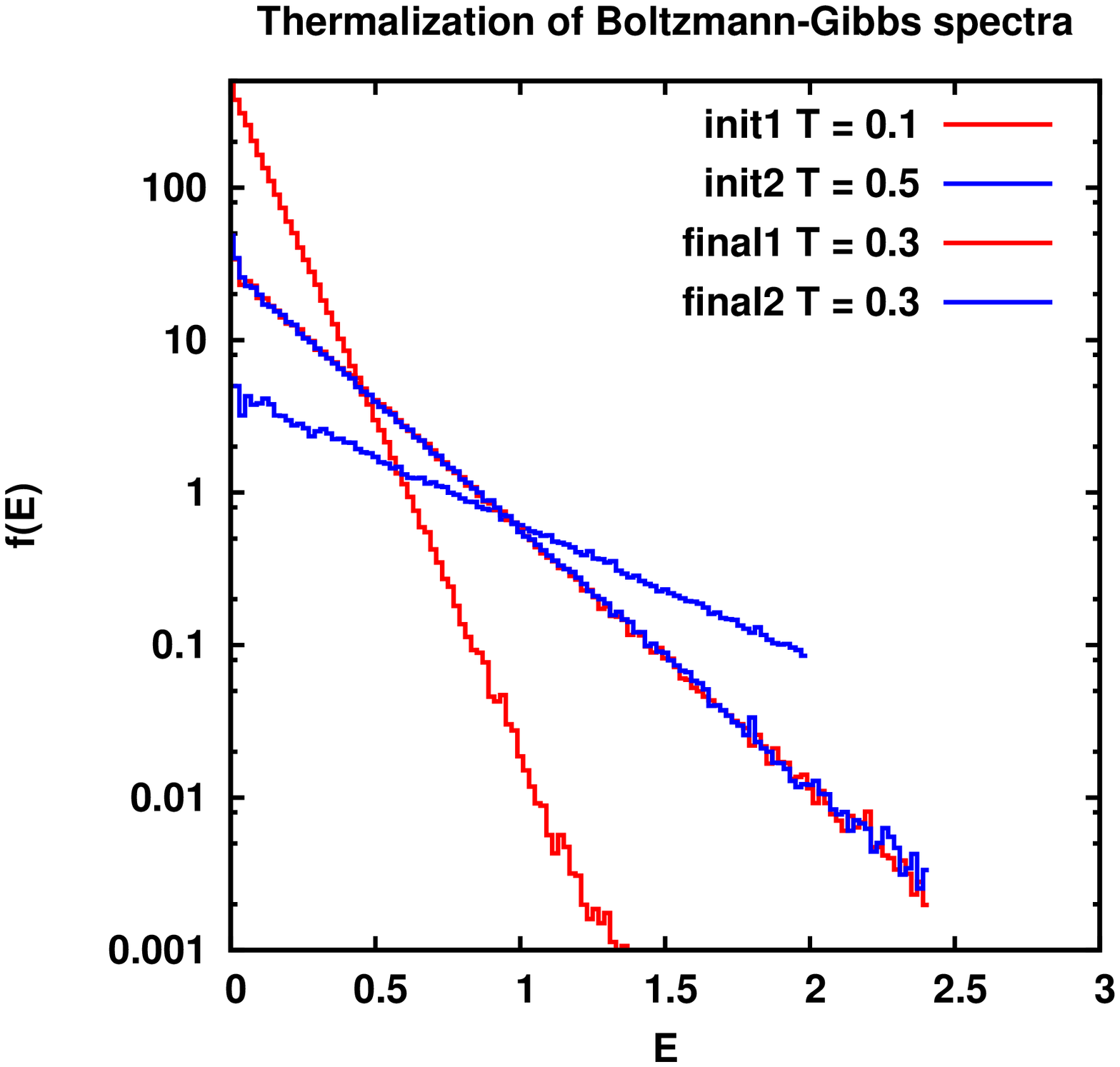}
 	\includegraphics[width=0.32\textwidth,angle=-90]{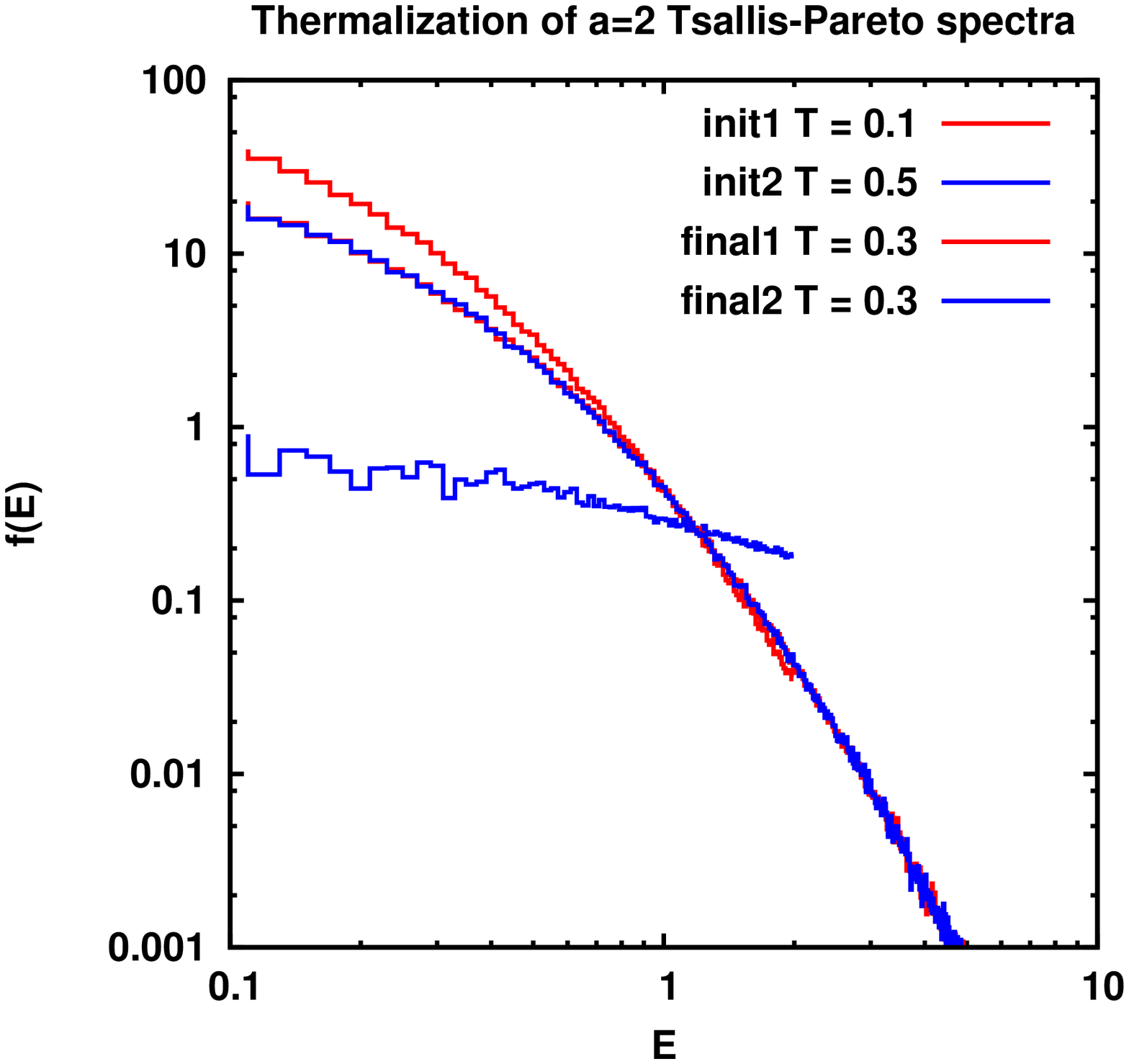}
\end{center}
\caption{Equilibration of two Boltzmann-Gibbs systems (\(a=0\), upper figure) \label{fig:BG} 
	plotted on a linear - logarithmic scale
	 and equilibration of two Tsallis-type non-extensive systems (\(a=2\), lower figure) \label{fig:TS}
	plotted on a double logarithmic scale. These are results for three-dimensional systems
	with $10$ collisions per particle on the average. Each subsystem consists of $100.000$ 
	massless particles.
}
\end{figure}

Our results show that the subsystems do equilibrate,  
they tend towards having a common stationary distribution.

We present examples with different initial conditions. 
We fix the particle numbers for each subsystem, \(N_1=N_2=100\,000\). 
The number of collisions in a typical simulation is \(N_{coll}=1\,000\,000\), 
so that \(N_{coll}/(N_1+N_2)=5\) collisions happen per particle. 
This quantity we use as an evolution parameter instead of the real time. 
This way we do not have to know differential cross sections; from the viewpoint of the fact
of equilibration its rapidity does not matter.

In the  figure \ref{fig:BG} energy distribution curves are shown:
the initial and the final ones and the ones after  $5$ collisions per
particle, respectively. The upper part plots a Boltzmann system (simulation with $a=0$)
the lower one a Tsallis system with the energy composition rule using $a=2$.
In the upper half a logarithmic -- linear plot is shown while in the lower half
a double logarithmic plot. These choices are selected by the respective high energy asymptotics;
exponential for a Boltzmann-Gibbs, while power-law for a Tsallis-Pareto distribution.
It is hard to distinguish the energy distributions in the subsystems in the final state,
the simulation curves are very close to each other. 
Therefore we conclude, that within numerical uncertainties a common stationary energy 
distribution is achieved.



\subsection{Equilibration of large subsystems}

Seeking for a canonical equilibrium state we have to maximize the total entropy given by
a general composition rule, $S(E_1,E_2)$, at the same time satisfying a constraint
which is in the general case also non-additive: $h(E_1,E_2)$ is constant.
For the moment we neglect the dependence on further thermodynamical variables;
usually the particle number $N$ and the volume $V$ is regarded to be proportional
and extensive.

In the traditional case both the entropy and the energy are combined
additively: $S(E_1,E_2)=S(E_1)+S(E_2)$ and $h(E_1,E_2)=E_1+E_2$. In the general case
by using corresponding formal logarithms the quantities $Y(S)$ and $X(E)$ have to be
considered as additive. Since for associative rules the formal logarithm is strict monotonic,
the maximum of the total entropy is achieved where $Y(S)$ has its extreme.
The general canonical principle is therefore given by
\be
 Y(S) - \beta X(E) = {\rm max.}
\ee{CANONICAL}
The parameter $\beta$ at this point is a Lagrange multiplier. Applying this for the equilibration
of two large subsystems, and assuming that the entropy of each systems depends only on its
own energy, one arrives at the equilibrium condition
\be
 \frac{Y'(S(E_1))}{X'(E_1)} \, S'(E_1) =
 \frac{Y'(S(E_2))}{X'(E_2)} \, S'(E_2) = \frac{1}{T}.
\ee{EQUIL_OF_TWO}
Comparing this with the general canonical form eq.(\ref{CANONICAL}) we obtain that $\beta=1/T$,
and $T$ is an absolute temperature in the classical thermodynamical sense.
Its relation to the entropy, however, has been generalized.
In particular for an additive entropy, but non-additive energy composition rule,
one arrives at $1/T=S'(E)/X'(E)$. The relation of this quantity to the logarithmic
spectral slope, $1/T_{{\rm slope}}=-d \ln f/dE = S'(E)$ leads to a practical tool for the
analysis of particle spectra in experiments. For the Pareto-Tsallis distribution
it is given by $T_{{\rm slope}}=T/X'(E)=T(1+aE)=T+(q-1)E$. The naive effort to extract a
temperature from energy spectra of particles, as it is a widespread usage in relativistic
heavy ion studies, only works if $q=1$, i.e. for spectra exponential in the particle energy.
Otherwise an energy dependent slope, and a curved spectrum in the logarithmic plot
has to be interpreted.

The inverse logarithmic slopes of single-particle kinetic energy spectra
in the generalized case are functions of the energy:
\be
 T_{{\rm slope}} = \frac{-1}{\pd{}{E} \ln f(E)}.
\ee{INVLOGSLOPE}
For the Tsallis-Pareto
distribution they are linear functions, $T_{{\rm slope}}=T+(q-1)E$.
Such slope parameters are plotted in Ref.\cite{PLAnoneq} 
for the respective subsystems
before and after equilibration ($10$ collisions per particle on the average).
Within numerical uncertainties it is clear that common-$a$ systems do equilibrate
at a common temperature also in the $a \ne 0$ case.

\begin{figure}
  \begin{center}
  \includegraphics[width=0.3\textwidth,angle=-90]{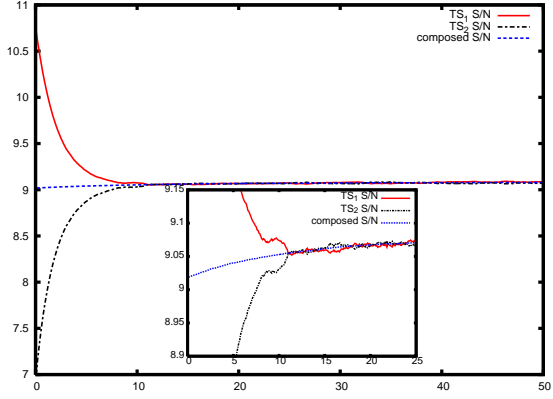}
  \end{center}
\caption{\label{ENTROPY_EQUIL}
   The evolution of the Boltzmann entropy per particle during collisions with non-additive
   energy composition rules: the hotter body cools, the cooler body warms up,
   while the total entropy also increases. In the insertion a magnification of the curves
   is shown.
}
\end{figure}

The (in our case Boltzmann) entropy also evolves due to the collisions.
In Fig.\ref{ENTROPY_EQUIL} the evolution of the entropy per particle is plotted
for the hot and cool subsystems, and for the total system respectively.
Since the composite system is combined from equal numbers of particles
in each subsystem, the total entropy per particle starts with the arithmetic mean
of the respective specific entropies. This value, however, rises somewhat,
featuring a trend according to the second law of thermodynamics.

\section{Power-law tailed hadron spectra, flow and quark coalescence}
\label{sec:4}

As an application of the above reviewed treatment of non-extensivity,
in this section we demonstrate that hadronic transverse momentum spectra
stemming from relativistic heavy ion collisions can be well described
by cut power-law spectra in statistical models. In order to do so, one
has to disentangle effects of a possible transverse flow on these
spectra and then test whether the result complies with the thermal assumption;
i.e. that the dependence on momenta is through a dependence on the kinetic
energy, $E-\mu=E-m$ only. The $\mu=m$ assumption corresponds to a vanishing
Fermi momentum for fermions, so this is the natural assumption at zero
net baryon density. Therefore transverse momentum spectra at mid-rapidity
are expected to follow such statistical model assumptions the best.

\subsection{Spectral temperatures in relativistic heavy ion collisions}

\vs
It has been long discussed, how a temperature can be conjectured from
observations on particle spectra produced in relativistic heavy ion collisions.
One intriguing way is to look at the transverse momentum, $p_T$, spectra around
mid-rapidity.
The different identified hadrons, mostly pions, kaons, protons and antiprotons,
have to demonstrate that their abundance in the momentum space depends on their kinetic energy;
this phenomenon at zero rapidity is the so-called $m_T-m$-scaling.
The transverse mass is given as $m_T=\sqrt{m^2+p_T^2}$, at strictly zero
rapidity this is the total relativistic energy.

\vs
The analysis is made a little more involved by the fact that the source emitting the
detected hadrons is flowing in all directions. 
The most prominent effects are due to a relativistic transverse flow
with velocity $v_T$ (and a corresponding Lorentz factor $\gamma_T=1/\sqrt{1-v_T^2}$
in units where $c=1$). The relativistic energy of a particle in the frame of the emitting
source cell is given by the J\"uttner variable: 
\be
E=u_{\mu}p^{\mu}=\gamma_T m_T \cosh(y-\eta)-\gamma_Tv_Tp_T\cos(\varphi-\Phi).
\ee{JUTTNER}
Here the four-velocity of the source and the actual four-momentum of the particle
are parametrized by rapidity and angle variables:
\ba
 u_{\mu} &=& (\gamma_T \cosh \eta, \gamma_T \sinh \eta, \gamma_Tv_T \cos \Phi,  \gamma_Tv_T \sin \Phi),
\nl
 p_{\mu} &=& (m_T \cosh y, m_T \sinh y, p_T \cos \varphi, p_T \sin \varphi). 
\ea{DEF_FOUR_VECTORS}
We consider a thermal model for the particle spectra; then the yield is supposed to
depend on the J\"uttner variable $E$ given by eq.(\ref{JUTTNER}).
Assuming a general distribution $f(E) \sim exp(-(X(E)-m)/T)$, which is monotonic decreasing, 
one finds its maximum at the minimum of $E$. This variable is minimal at the rapidity
$y_{{\rm min}}=\eta$, and angle $\varphi_{{\rm min}}=\Phi$, giving 
\be
  E_{{\rm min}}=\gamma_Tm_T-\gamma_Tv_Tp_T. 
\ee{MIN_TRV_ENERGY}
This Lorentz-boosted transverse energy reaches its minimum at the 
transverse momentum value \hbox{$p_{T,{\rm min}}=m\gamma_Tv_T$,}
leading to \hbox{$m_{T,{\rm min}}=m\gamma_T$} and \hbox{$E_{{\rm min}}=m$.} 
The expansion around this minimum in the $p_T$-distribution is an effective Gaussian:
\be
 e^{-(E-m)/T} \approx \exp\left({-\frac{(p_T-m\gamma_Tv_T)^2}{2m\gamma_T \, T\gamma_T}}\right).
\ee{EFF_GAUSS}

\begin{figure}
\centerline{\includegraphics[width=0.3\textwidth]{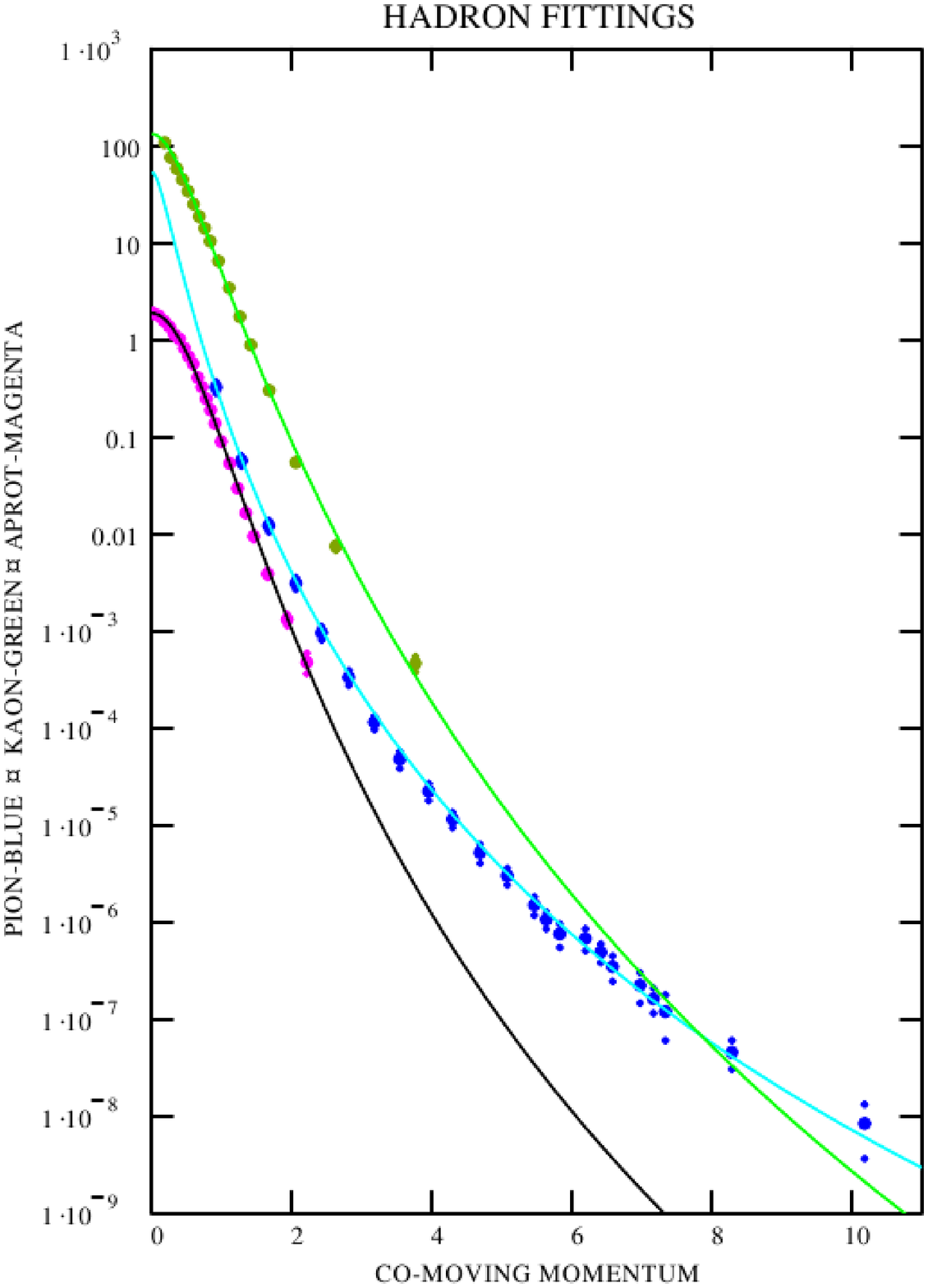}}
\centerline{\includegraphics[width=0.3\textwidth]{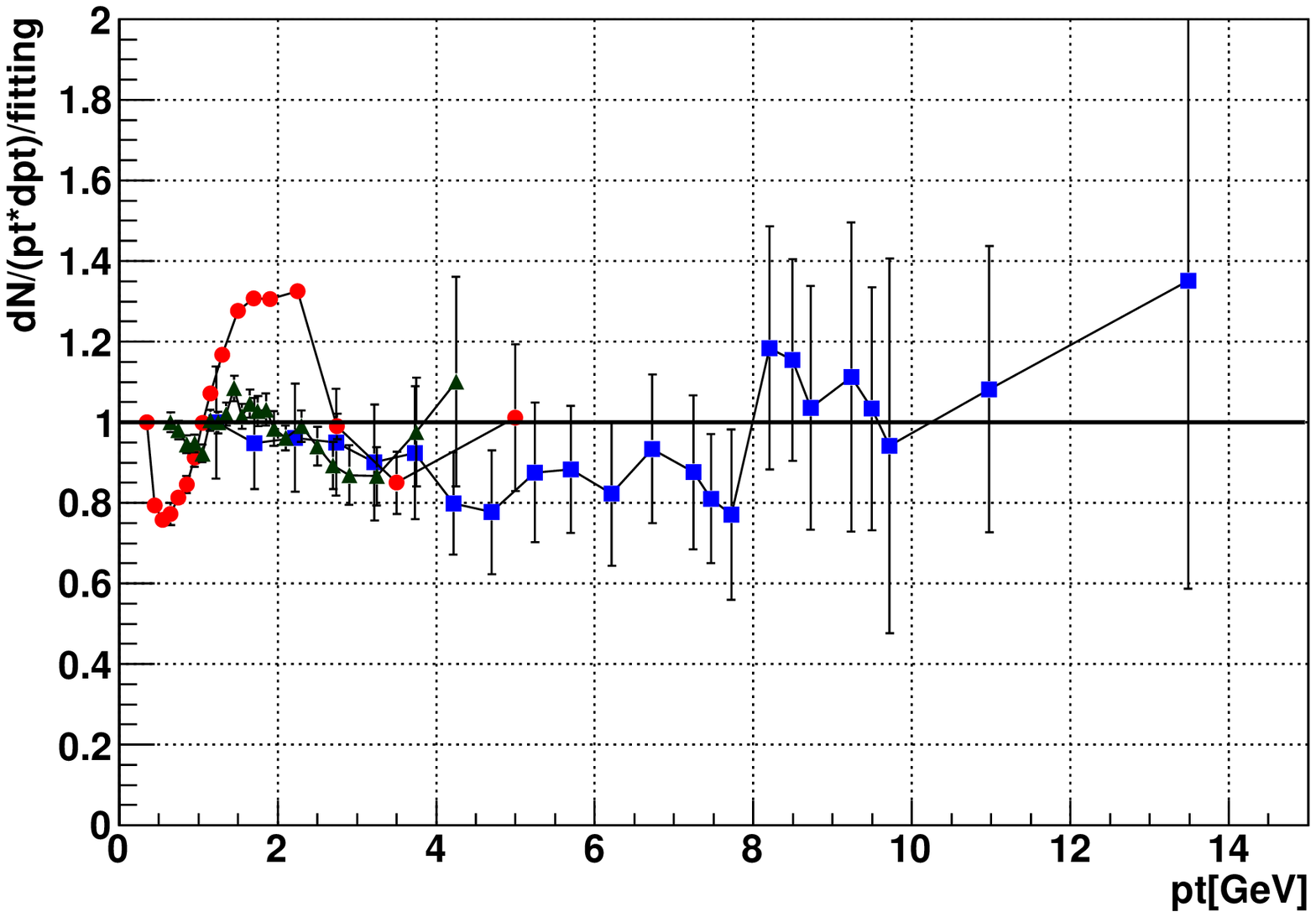}}
\caption{\label{PiKpfits}
  General shape of $p_T$ spectra for pions, kaons and antiprotons
  in relativistic heavy ion experiments (upper figure). 
  A fit is done by using
  for $X(E)$ the Tsallis-Pareto form with parameters $T$ and $a$, corresponding to
  a common temperature of $T(m_i)= 0.160$ MeV  for the different particles.
  and a transverse flow velocity $v_T=0.52$.
  In the lower part the ratio of the Tsallis fit to the experimental values
  can be inspected in a linear plot. 
}
\end{figure}

In fact, according to experimental findings at RHIC the observed particle spectra have to be
corrected for a transverse flow in order to reach $m_T$-scaling.

\subsection{Non-extensivity in quark matter and in hadron matter}

\begin{figure}
  \begin{center}
  {\includegraphics[width=0.35\textwidth]{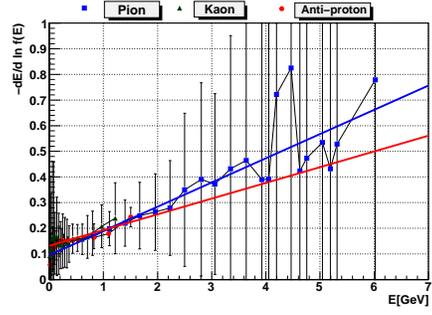}}
 \end{center}
\caption{\label{SLOPES}
    Inverse logarithmic slopes, $T_{{\rm slope}}(E)=-dE/d\ln f(E)$ 
    extracted by numerical derivation from the experimental hadronic spectra 
    (after subtracting a common flow effect).
    The full lines correspond to a common meson and baryon
    fit their steepness keeping the ratio 2:3 predicted by the quark coalescence picture.
}
\end{figure}

We conjecture that the power-law tails observed in hadronic spectra may stem from 
non-extensivity of the suddenly hadronizing quark matter.
We look for a connection between quark and hadron spectra in the framework of the quark coalescence model.
A coalescence of  a quark and an antiquark into a meson produces
a yield proportional to the quantity:
\be
 F(\vec{p}) = \int f\left(E(\vec{P}/2+\vec{q})\right) f\left(E(\vec{P}/2-\vec{q})\right) C(\vec{q}) \, d^3q. 
\ee{COALESCENCE}
Here we integrate over the relative momentum of the quarks with a coalescence
factor, $C(\vec{q})$, for which a simple model has been utilized \cite{JPG.COAL}.
For common momenta much larger than the relative one $|\vec{P}|\gg |\vec{q}|$
on obtains
\be
 F(\vec{P}) \approx \,  f^2\left(E(\vec{P}/2)\right) \int  C(\vec{q}) \, d^3q. 
\ee{LARGE_MOM_COAL}
In particular light hadrons made from massless quarks follow the quark-scaling rule:
\be
 f_{{\rm hadron}}(E) \propto f^n(E/n).
\ee{QUARK_RULE}
As a consequence particular properties of the non-extensive thermal model between quark and
hadron matter also scale:
$T_{{\rm mesons}} = T_{{\rm baryons}} = T_{{\rm quarks}}$ for the temperature, while 
  $q_{{\rm mesons}}-1 = (q_{{\rm quarks}}-1)/2$ for mesons and 
  $q_{{\rm baryons}}-1 = (q_{{\rm quarks}}-1)/3$ for baryons.
Since for a Tsallis-Pareto distribution the inverse logarithmic slope turns out to be
\be
 T_{{\rm slope}} = T + (q-1)(E-m),
\ee{TSALLISINVLOGSLOPE} 
the rise of these slopes reflect the non-extensivity parameters.
In Fig.\ref{SLOPES} we show the test of the coalescence model prediction for
the meson to baryon ratio.

\begin{figure}
\centerline{\includegraphics[width=0.25\textwidth,angle=-90]{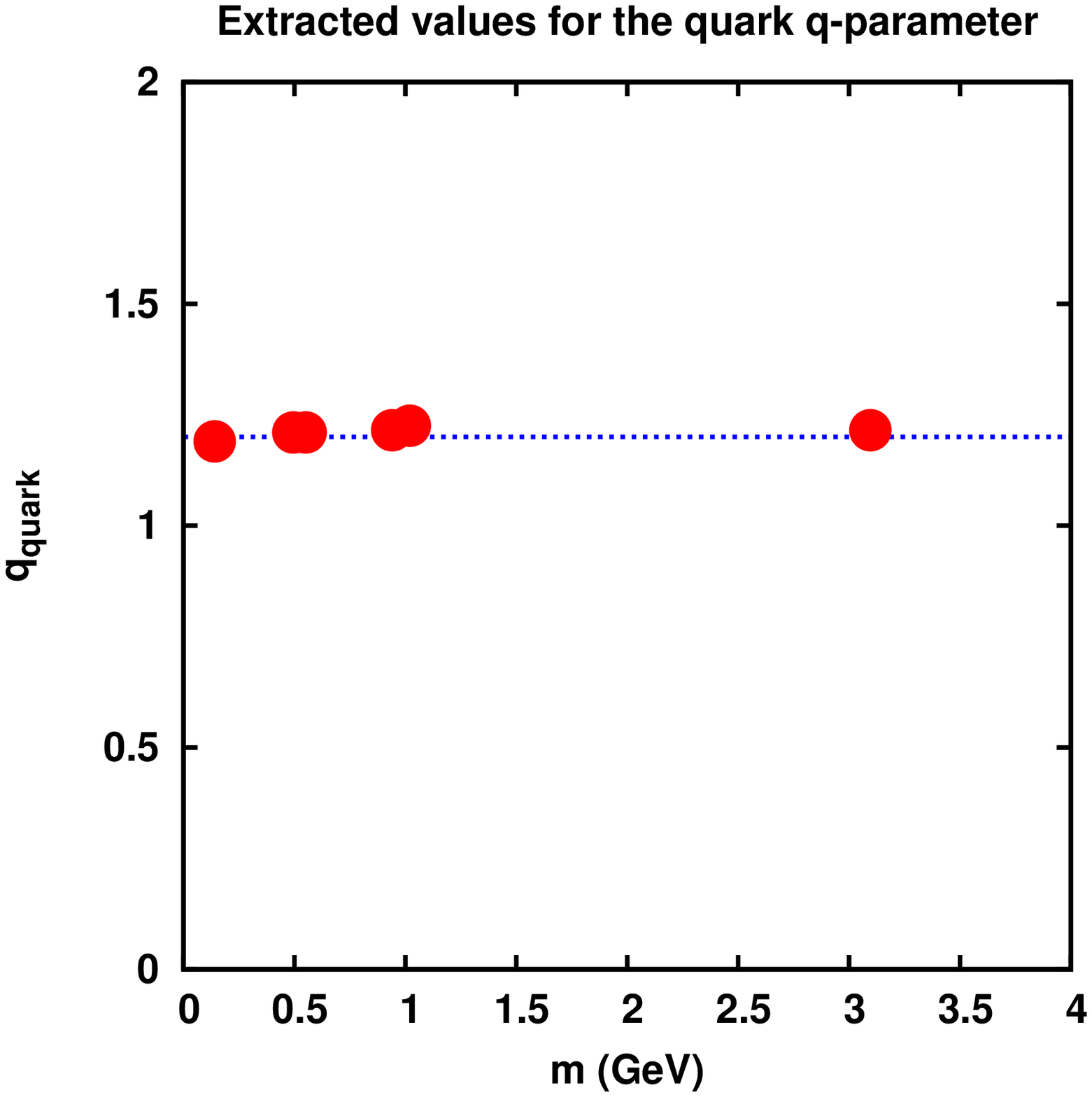}}
\centerline{  \includegraphics[width=0.25\textwidth,angle=-90]{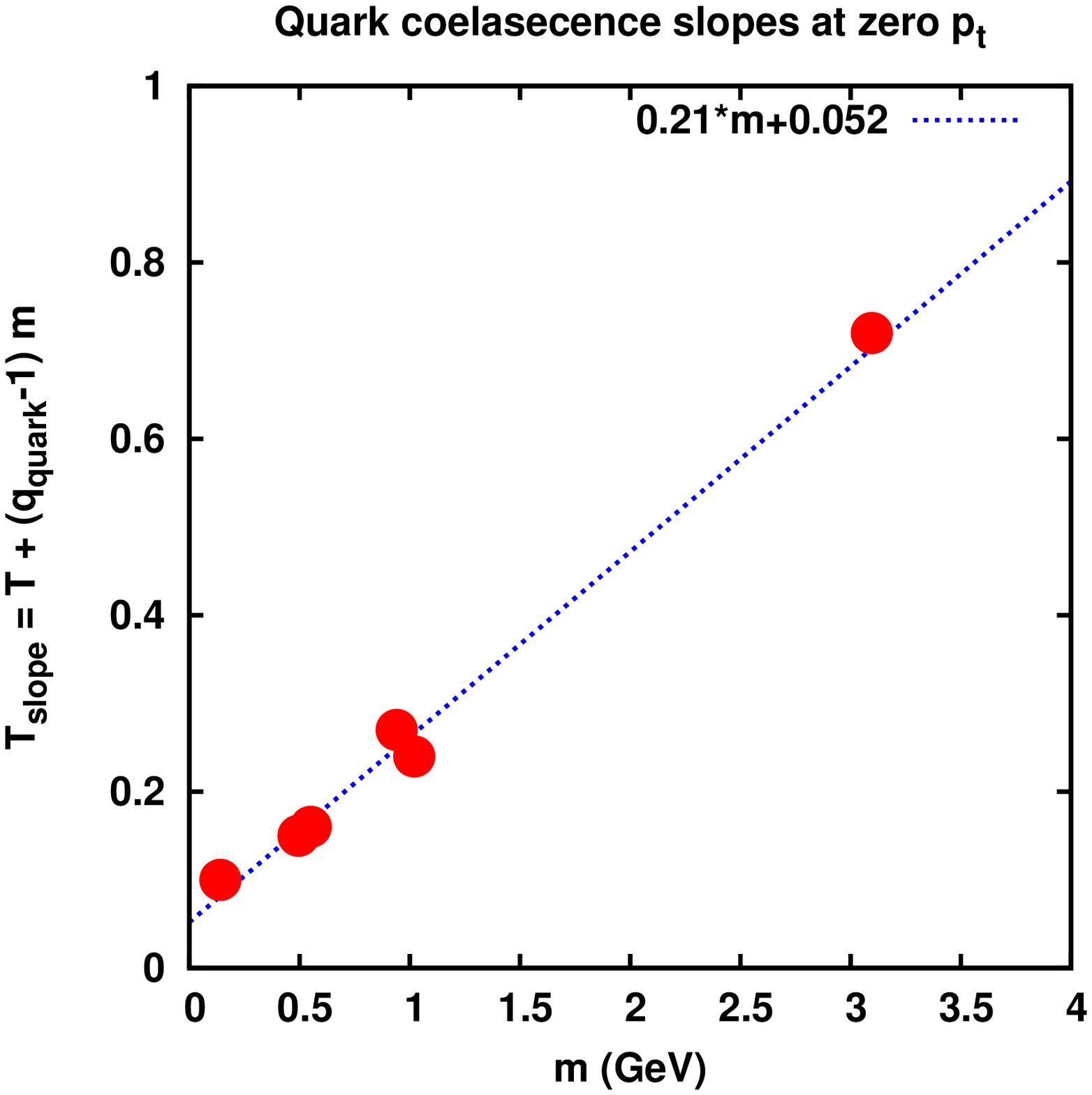} }
\caption{\label{QuarkMatterQ}
  The $q$ parameter of quark matter extracted from hadronic spectra assuming
  quark coalescence at a sudden hadron formation (upper picture).
  The spectral inverse slope as a function of the minimal energy $E_{{\rm min}}=m$ agree with
  the linear prediction from the coalescence scaling.
}
\end{figure}

Furthermore these predictions of the non-extensive phenomenology 
meet the curves from pQCD calculations, with the following
surmised properties of quark matter at RHIC: $T=140\ldots 180$ MeV, $q=1.22$, $v_T=0.6$. 
\cite{SQM2007}. Deviations from the product rule suggested by the simplest
quark coalescence idea occur at $p_T$ values lower than $1$ GeV.
We attribute these to a further constituent in real hadrons, namely
non-perturbative gluons simulated by a string energy contribution\cite{SQM2008}.

We note that a stringy interaction remainder above the color deconfinement
temperature $T_c \approx 170$ MeV in quark gluon plasma also describes the
main effects on the quark matter equation of state seen in lattice QCD calculations
successfully\cite{BiroCleymans}. Both the presence of a string like pair potential
for a however minor percentage of pairs,
as a microscopic model, and the assumption of non-extensivity, $q > 1$,
as a descriptive phenomenology are able to explain the value of the energy
per particle, $E/N=6T=1$ GeV, which has been found by fitting 
statistical hadronic resonance gas models with Boltzmann distributions.

Finally some remarks are in order to the Tsallis-Pareto fits to
energy spectra. In several cases naively a fit is done in the original form:
\be
 f(E) \sim \left(1+(q-1)\frac{E}{T} \right)^{-\frac{1}{q-1}}
\ee{NAIVECUTFIT}
to which the following inverse logarithmic slope dependence belongs:
\be
 T_{{\rm slope}} = T + (q-1)E.
\ee{NAIVESLOPE}
A more sophisticated approach (as one suggested in \cite{Lavagnoetal}), however,
uses the original Tsallis-Pareto form for the number density distributions of
particles and for the generating thermodynamical potential, for the logarithm
of the canonical partition function. This way in this second approach the
energy distribution is described by the $q$-th power of the naive factor:
\be
 f(E) \sim  \left(1+(\tilde{q}-1)\frac{E}{\tilde{T}} \right)^{-\frac{\tilde{q}}{\tilde{q}-1}} 
\ee{QCUTFIT}
and the corresponding inverse slope
\be
 T_{{\rm slope}} = \frac{1}{\tilde{q}} \tilde{T} + \left(1-\frac{1}{\tilde{q}} \right)E.
\ee{QSLOPE}
We observe that the qualitative behavior is the same, but the interpretation of the fit
parameter is different in these different approaches. The correspondence between the
energy spectrum fit parameters is given as
\ba
 T &=& \tilde{T}/\tilde{q}, \nl
 q &=& 2 - 1/\tilde{q}.
\ea{TWOFITS}
In a sense $q$ and $\tilde{q}$ are double-duals of each other, both using the $1/q$- and
the $2-q$-duality. Also the estimated temperature parameter differ.
Typical values from relativistic heavy ion experiments are $q\approx 1.2$ and $\tilde{q}\approx 1.25$,
as well as, $\tilde{T}=1.25T$.

\section{Conclusion}

In conclusion we reviewed basic concepts of non-extensive thermodynamics
which may be relevant in understanding particular features of hadronic
spectra stemming from relativistic heavy ion collisions. The overall presence of
relativistic speeds of particles in the physical system under investigation
on the other hand offers a unique possibility to study and - whenever necessary -
to generalize familiar thermodynamics.

We presented some general arguments for a possible need to face with
total energy and entropy not being proportional to the particle number
even in the large $N$ limit. These arguments are based on the long range nature of pair
interactions. This phenomenon, called non-extensivity, was then related to the
generalization of composition rules of the familiar thermodynamical extensives,
like energy and entropy. We have mathematically proved that abstract composition
rules become symmetric and associative in the large $N$ limit, provided that
the composition function, $h(x,y)$, is at least right-sided differentiable at $y=0^+$.
This means that associative composition rules constitute attractors among
all rules when approaching the thermodynamical limit. As a consequence the
associativity of the composition rule is a thermodynamical requirement.

The key quantity in this proof, the formal logarithm, relates the abstract
composition rule to the addition of the system size indicator, to the particle
number $N$. We gave the formula how to construct it. Based on the formal
logarithm the widely used {\it deformed} exponential and logarithm functions
can easily be derived. While the former describes the energy distribution in
canonical equilibrium, the latter defines a generalized formula for the entropy.
An additive entropy can be always gained from this expression by taking its
formal logarithm.

We presented some often used composition rules together with the corresponding
equilibrium energy distributions and entropy formulas including the traditional
Boltzmann-Gibbs formula derived from the simple addition (extensivity),
the Tsallis rule, the Kaniadakis rule and - for the sake of demonstration -
the Einstein rule for composing relativistic velocities. Our general method
in this case leads to the rapidity as the additive formal logarithm.
Among non-associative composition rules the class of $h(x,y)=x+y+G(xy)$ is
found to be particularly interesting in high energy physics, since it
asymptotically approaches the Tsallis rule leading to power-law tailed
energy distributions in canonical equilibrium. We have demonstrated that
such a composition rule may emerge in the extreme relativistic kinematics
limit from an energy correction to a pair of particles in a medium 
which is a function of the Lorentz invariant relative momentum squared
variable $Q^2$. In fact this is frequently the case when following several
interactions among partons according to the formulas derived from (or at least
motivated by) QCD. Finally it is interesting to note that the elementary property,
$h(x,0)=x$ is related to $\sigma(1)=0$ property of the entropy density function
if the composition rule $h(x,y)$ is assumed for composite states with
factorizing probabilities. In this case the general result $\sigma(p)=\ln_a(1/p)$
emerges in the thermodynamical limit, with $\ln_a=L^{-1}\circ \ln$ being the
corresponding deformed logarithm function.

The - in some sense opposite - requirement, i.e. aiming at an additive
entropy formula while the probabilities do not factorize, but their 
logarithms follow a general composition rule $h(x,y)$ instead of the
usual addition, leads to a more complex relation between the formal
logarithm of the rule, $L$, and the entropy density $\sigma$.

In the second part of this review we compiled the most important
numerical results on parton cascade simulations of non-extensive
systems. Following the presentation of a class of generalized
Boltzmann equations, and proving that the second law of thermodynamics
can only be fulfilled if the derivative of the entropy density 
is a linear expression of the deformed logarithm of the one-particle
phase space density (cf. eq.(\ref{SMALL_SIGMA_PRIME})), we presented
some details of the kinematical description of relativistic
particles in such simulations. We payed special attention to the
random choice of particle momenta after an, in energy non-additive,
pair collision (which can have a physical reason in the influence
of third or further particles, or fields in a dense medium).

Results on the phase space evolution under non-exten\-sive energy
composition rules were presented demonstrating the ability of
such a computer simulation to generate power-law tailed energy spectra
in the detailed balance state of the non-extensive Boltzmann equation.
The important question of equilibration between two large
subsystems, related to the zeroth theorem of thermodynamics,
was also investigated by us numerically in this framework.
We found that non-extensive systems with the power-law tailed
Tsallis-Pareto energy distributions do behave as they should,
just the thermodynamic temperature, $T$ is related to the
microcanonical equation of state, $S(E)$, by receiving corrections
due to the formal logarithms of the entropy and energy composition
rules (cf. eq.(\ref{EQUIL_OF_TWO})). 

Finally our studies on the hadronization of quark matter in
relativistic heavy ion collisions revealed that if the quark coalescence
is a dominant mechanism, then the non-extensivity parameter,
$q-1=aT$ also must show the quark number scaling. This assumption
can be and should be tested on experimental data and should be
related to other information on quark coalescence, e.g. to those
obtained from studies of the elliptic flow.

Certainly there remain open questions for further research.
Among them the study of the quark matter equation of state with
elementary field theory means, as lattice QCD, in a non-extensive
canonical state is still a hard challenge. Also the determination
of the pair correlation function, $g(r)$, from first principles
in microscopical calculations should help to identify those
physical situations where the concepts  and formalism of non-extensive
thermodynamics have to be used. Meanwhile the physical reason
for a non-exponential energy distribution can be numerous.
The quark gluon plasma is a wonderful candidate for finding
non-extensive behavior, since long range effects are there at
any finite temperature.

%

\vspace{3mm}
{\bf Acknowledgments}

This work has been supported by the Hungarian National Science Fund, OTKA
(K49466, K68108). Discussions with
C.~Tsallis,  G.~Wilk, T.~Kodama, G.~Kaniadakis, P.~V\'an and A.~L\'aszl\'o
are gratefully acknowledged.





%
%
%
%

\newcommand{\etal}{{\it et.al.}}

\end{document}